\documentclass{article}

\usepackage{microtype}
\usepackage{graphicx}
\usepackage{subfigure}
\usepackage{booktabs} 

\usepackage{pgfplots}
\usepackage{amsmath}
\usepackage{amsfonts}
\usepackage{multirow}
\usepackage{makecell}
\usepackage{enumerate}

\usepackage{hyperref}

\usepgfplotslibrary{groupplots}
\usetikzlibrary{pgfplots.statistics}
\pgfplotsset{width=10cm,compat=1.9}

\usepackage[accepted]{main_arxiv}

\newcommand{\system}{Cephalo\ }
\newcommand{\systemns}{Cephalo}

\begin{document}

\twocolumn[
\mlsystitle{\systemns: Harnessing Heterogeneous GPU Clusters \\for Training Transformer Models}

\mlsyssetsymbol{equal}{*}

\begin{mlsysauthorlist}
\mlsysauthor{Runsheng Benson Guo}{waterloo}
\mlsysauthor{Utkarsh Anand}{waterloo}
\mlsysauthor{Arthur Chen}{waterloo}
\mlsysauthor{Khuzaima Daudjee}{waterloo}
\end{mlsysauthorlist}

\mlsysaffiliation{waterloo}{School of Computer Science, University of Waterloo, Waterloo, Canada}

\mlsyscorrespondingauthor{Runsheng Benson Guo}{r9guo@uwaterloo.ca}

\mlsyskeywords{Machine Learning, MLSys}

\vskip 0.15in

\begin{abstract}
    Training transformer models requires substantial GPU compute and memory resources. In homogeneous clusters, distributed strategies allocate resources evenly, but this approach is inefficient for heterogeneous clusters, where GPUs differ in power and memory. As high-end GPUs are costly and limited in availability, heterogeneous clusters with diverse GPU types are becoming more common. Existing methods attempt to balance compute across GPUs based on capacity but often underutilize compute due to memory constraints. We present \systemns, a system that optimizes compute and memory usage by decoupling compute distribution from training state assignment. 
    \system outperforms state-of-the-art methods by achieving significantly higher training throughput while supporting larger models and batch sizes.
\end{abstract}
]



\printAffiliationsAndNotice{}  

Transformer models \cite{vaswani2017attention} have demonstrated state-of-the-art performance 
in many domains including natural language processing (NLP), computer vision, and recommendation systems \cite{devlin2018bert, vit,sun2019bert4rec}. 
In particular, large language models (LLMs), which are based on the transformer architecture, have significantly
advanced NLP tasks such as question-answering, translation, and summarization \cite{devlin2018bert, gpt, pegasus}.
Since increasing model size can yield significant 
improvements in accuracy, this has led to the development of larger models that often 
exceed modern GPU compute and memory capabilities \cite{pati2023computation}. 

Consequently, many strategies have been proposed to distribute and parallelize training across multiple GPUs. 
Data parallelism replicates the model
across GPUs, each training on a different subset of the inputs in parallel. 
Model parallelism splits the model across GPUs, with each GPU storing and processing only a partition of the model's parameters.

\begin{figure}
  \centering
  \begin{tikzpicture}
    \begin{axis}[
      xlabel={Hour},
      ylabel={Available GPUs},
      grid=major,
      legend style={at={(1.05,0.5)}, anchor=west, font=\tiny}, 
      xtick={0,4,...,24},
      ymin=-2, ymax=35,
      width=0.8\linewidth,
      height=3cm,
      tick label style={font=\small}, 
      label style={font=\small}, 
      title style={font=\small}, 
    ]
      \addplot[color=cyan, mark=o, mark size=1.5pt] coordinates {
        (1,32) (2,32) (3,32) (4,32) (5,32) (6,32) (7,32) (8,32) (9,32) (10,32) (11,32) (12,32)
      }; 
      \addplot[color=blue, mark=square*, mark size=1.5pt] coordinates {
        (1,8) (2,16) (3,16) (4,24) (5,0) (6,32) (7,32) (8,16) (9,16) (10,16) (11,32) (12,32)
      }; 
      \addplot[color=red, mark=triangle*, mark size=1.5pt] coordinates {
        (1,16) (2,32) (3,24) (4,8) (5,0) (6,32) (7,8) (8,0) (9,32) (10,0) (11,32) (12,0)
      }; 
      \addplot[color=orange, mark=diamond*, mark size=1.5pt] coordinates {
        (1,0) (2,8) (3,0) (4,24) (5,8) (6,8) (7,0) (8,0) (9,0) (10,8) (11,0) (12,8)
      }; 
      \addplot[color=black, mark=x, mark size=1.5pt] coordinates {
        (1,0) (2,0) (3,0) (4,0) (5,0) (6,0) (7,0) (8,0) (9,0) (10,0) (11,0) (12,0)
      }; 
      \legend{T4, V100, A10G, A100, H100}
    \end{axis}
  \end{tikzpicture}
  \caption{Hourly AWS GPU availability over 12-hour period.}
  \label{fig:gpu_availability}
\end{figure}
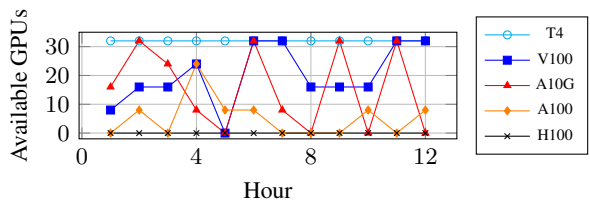

While existing parallelization strategies typically assume GPU homogeneity, ML practitioners, in reality, 
often do not have sufficiently large homogeneous 
clusters for training transformers \cite{park2020hetpipe,miao2023sdpipe}. 
For example, a small-scale company or research lab may not have the resources to purchase an entire cluster 
of the latest GPUs. 
Instead, they are more likely to accumulate a diverse array of GPUs with varying compute and memory capacities 
over time \cite{partial_reduce,flashflex, metis}. Cloud platforms like AWS offer VMs with a variety of GPU
models, but due to high demand, each model is available only in limited quantities. Figure \ref{fig:gpu_availability} plots a trace of GPU availability
on AWS over a 12-hour period in the us-west region. High-end GPUs (A100, H100) are almost always unavailable, and even
mid-tier GPUs (A10G, V100, T4) are limited due to capacity and quota. Thus, it is challenging to reserve a large homogeneous cluster 
of GPUs.

By assembling \textit{heterogeneous} clusters with different GPU models, users 
can leverage a larger pool of compute resources for training.
However, existing systems are unable to utilize resources efficiently in heterogeneous clusters. 
Systems for homogeneous clusters divide compute and memory demands
evenly among all GPUs \cite{galvatron, zero3, shoeybi2019megatron}.
In clusters with varying GPU capabilities, training is bottlenecked by the slowest GPU, leaving 
faster GPUs idle. 
Additionally, training fails if GPUs with the lowest memory run out, even if others have unutilized memory.

Heterogeneity-aware training methods have been proposed, which aim to balance computational load across GPUs.
For instance, in data parallelism, the batch of inputs is distributed unevenly across GPUs according to 
their relative computational speeds \cite{moreno2020training,kim2022scale,whale}. Systems using model parallelism partition 
the model's layers or parameters unevenly across GPUs to balance computation \cite{pipedream,hap}. 
Recent methods integrate both data and model parallelism to further optimize compute distribution \cite{flashflex,metis}.

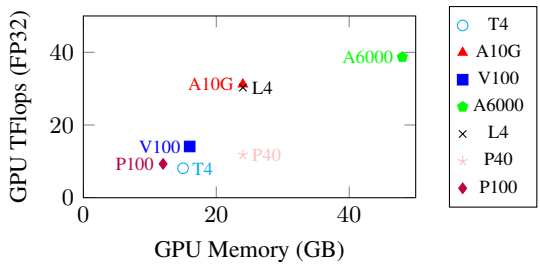
\begin{figure}[htbp]
  \centering
  \begin{tikzpicture}
  \begin{axis}[
      xlabel={GPU Memory (GB)},
      ylabel={GPU TFlops (FP32)},
      xmin=0, xmax=50,
      ymin=0, ymax=50,
      legend pos=north west,
      grid style=dashed,
      width=6cm, 
      height=4cm, 
      tick label style={font=\small}, 
      label style={font=\small}, 
      title style={font=\small}, 
      legend style={
          at={(1.1, 0.5)}, 
          anchor=west,    
          font=\scriptsize,
          /tikz/every even column/.append style={column sep=0.5cm}
      }
  ]
  \addplot[
    only marks,
    mark=o, 
    mark size=2pt,
    color=cyan,
    every node near coord/.append style={font=\tiny, anchor=west},
    ] coordinates {(15.00, 8.10)} node[anchor=west] {\scriptsize T4};

  \addplot[
    only marks,
    mark=triangle*, 
    mark size=2pt,
    color=red,
    every node near coord/.append style={font=\tiny, anchor=west},
    ] coordinates {(24.00, 31.20)} node[anchor=east] {\scriptsize A10G};

  \addplot[
    only marks,
    mark=square*, 
    mark size=2pt,
    color=blue,
    every node near coord/.append style={font=\tiny, anchor=west},
    ] coordinates {(16.00, 14.10)} node[anchor=east] {\scriptsize V100};
  
  \addplot[
    only marks,
    mark=pentagon*, 
    mark size=2pt,
    color=green,
    every node near coord/.append style={font=\tiny, anchor=west},
    ] coordinates {(47.99, 38.7)} node[anchor=east] {\scriptsize A6000};

  \addplot[
    only marks,
    mark=x, 
    mark size=2pt,
    color=black,
    every node near coord/.append style={font=\tiny, anchor=west},
    ] coordinates {(24.00, 30.3)} node[anchor=west] {\scriptsize L4};

  \addplot[
    only marks,
    mark=star, 
    mark size=2pt,
    color=pink,
    every node near coord/.append style={font=\tiny, anchor=west},
    ] coordinates {(24.00, 11.8)} node[anchor=west] {\scriptsize P40};

  \addplot[
    only marks,
    mark=diamond*, 
    mark size=2pt,
    color=purple,
    every node near coord/.append style={font=\tiny, anchor=west},
    ] coordinates {(12.00, 9.30)} node[anchor=east] {\scriptsize P100};
  \legend{
    T4,
    A10G,
    V100,
    A6000,
    L4,
    P40,
    P100
  }
  \end{axis}
  \end{tikzpicture}
  \caption{GPU TFlops (FP32) vs. Memory Capacity.}
  \label{fig:gpu_latency_vs_memory}
\end{figure}

  These load-balancing techniques allocate memory on each GPU proportional to its computational capacity. 
  In data parallelism, a GPU assigned a larger batch of inputs requires more memory for operations and activation 
  storage. Similarly, in model parallelism, a GPU handling a larger model shard demands additional memory to 
  maintain the training state. However, as shown in Figure \ref{fig:gpu_latency_vs_memory}, 
  a GPU's memory capacity does not always scale with its compute speed. 
  This mismatch can prevent effective computational load balancing due to memory limitations. 
  For example, while the L4 GPU offers significantly faster computation than the P40, both GPUs have the 
  same memory capacity, meaning the L4 may lack sufficient memory to handle twice the computational workload.

Thus, existing systems are susceptible to both: (i)
underutilizing compute on GPUs with low memory capacity relative to compute speed, and (ii) underutilizing 
memory on GPUs with high memory capacity relative to compute speed.

In light of these shortcomings, we designed \systemns, a system capable of effectively utilizing
the aggregate compute \textit{\textbf{and}} memory resources in heterogeneous GPU clusters when training transformer models.

\system partitions the global batch of training inputs
unevenly across GPUs to control the computational workload assigned to each GPU.
To control the memory utilization on each GPU, \system combines the following strategies:

\begin{enumerate}[(i)]
    \item The training state (parameters, gradients, and optimizer state) is sharded across the GPUs to
balance memory utilization. Each GPU can store anywhere from none of the training state to the entire training state.
Flexibly sharding the training state is implemented on top of \textit{Fully Sharded Data Parallelism} (FSDP) \cite{fsdp}, which evenly distributes the training state across GPUs.

\item Gradients can be accumulated over multiple smaller batches to replicate training on larger batch sizes while using less
memory for compute operations.

\item Memory for storing intermediate activation values is eliminated with a combination of recomputing and offloading activations
to CPU when they are not used. 
\end{enumerate}

These mechanisms used for controlling computational workload and memory can be applied independently.
This allows
\system to decouple the assignment of compute and memory to each GPU and fully 
utilize the aggregate GPU compute and memory available within a heterogeneous cluster of GPUs in scenarios where 
state-of-the-art systems fall short. 

In this paper, we make the following contributions:
\begin{enumerate}
  \item We designed and implemented \systemns, a system for training transformer models on heterogeneous GPU clusters that jointly optimizes compute and memory distribution to maximize training throughput by efficiently utilizing resources across GPUs. \system includes an optimizer to divide training data, manage training state, and configure gradient accumulation to accommodate resource heterogeneity.
  \item We integrate gradient accumulation and activation offloading efficiently in FSDP. Our implementation
  of gradient accumulation minimizes the overhead of gathering training state. Our activation offloading reduces memory usage from gradient accumulation and overlaps with compute to hide transfer latency.
  \item We perform an extensive evaluation of \system on heterogeneous GPU clusters with up to $64$ GPUs
  and on transformer models with up to $7$ billion parameters.
  We show that \system is able to achieve up to $10\times$ higher training throughput
  than comparative state-of-the-art heterogeneous training systems 
  while supporting training for larger models
  and batch sizes.
\end{enumerate}

\section{Background and Related Work}
\subsection{Training Transformers}
Transformer models consists of a sequence of identical encoder
and decoder layers \cite{vaswani2017attention}, containing computationally expensive self-attention mechanisms and feed-forward networks. 
Training with optimizers like Adam \cite{kingma2014adam} requires 16 bytes of memory per model parameter on the GPU \cite{zero3, smith2022using}, covering not only the model parameters but also their gradients and optimizer state.
Besides maintaining the training state, GPU memory is also required to run operations and
store intermediate activation outputs. 
Thus, even a mid-sized transformer like Llama 7B \cite{touvron2023llama} requires more memory for training than the 80GB available on cutting-edge H100 GPUs.

\subsection{Distributed Training}
Given these substantial GPU memory and computational requirements, 
transformer training is typically parallelized.

\textbf{\textit{Data Parallelism}} \cite{sergeev2018horovod} replicates the model 
across GPUs, each computing a gradient on its own batch of data. 
This ``vanilla" data parallelism works only if each GPU can store the entire training state. 
ZeRO-3 \cite{zero3} is a variant of data parallelism that 
evenly shards the training state across GPUs. This allows for larger models to be trained by reducing 
the training state stored per GPU by a factor of $N$, albeit at the 
cost of $50\%$ more communication. Fully sharded data parallelism (FSDP) \cite{fsdp} is an efficient 
implementation of ZeRO-3 in PyTorch \cite{li2020pytorch}. 

\textbf{\textit{Model Parallelism}} 
partitions a model across GPUs, with each GPU storing only the training state for its assigned shard, enabling the training of models larger than a single GPU's memory. Pipeline parallelism \cite{huang2019gpipe, narayanan2019pipedream} divides the model into stages of consecutive layers, passing activations and gradients between stages. It parallelizes compute by processing microbatches in a pipeline across these stages. 
Tensor parallelism \cite{shoeybi2019megatron, shazeer2018mesh} is another form of model parallelism that distributes inputs, computation, and parameters for each layer evenly across GPUs, with all-to-all communication reassembling outputs between layers.

\subsection{Heterogeneous GPU Clusters}
Many distributed training systems assume a homogeneous GPU cluster, dividing compute and memory demands equally. However, most organizations lack large homogeneous clusters due to frequent GPU release cycles, high upgrade costs, GPU shortages, and limited cloud availability \cite{partial_reduce, datanami2023awscompute, jayaram2023sia, cross-region}
As a result, organizations often rely on clusters with GPUs
from different generations, 
which offer substantial compute power in aggregate. 
Thus, training on heterogeneous clusters has gained attention as it allows organizations to 
leverage all available GPU resources for training \cite{park2020hetpipe, metis, whale, hap, flashflex}.

\subsection{Heterogeneous Training}
Existing systems for training typically assume a cluster of homogeneous GPUs and split the 
workload evenly across GPUs. This strategy is susceptible to underutilizing GPU 
resources on a heterogeneous cluster since faster GPUs will be idle while waiting to synchronize 
with slower GPUs.

Systems like Whale \cite{whale, moreno2020training} propose to mitigate bottlenecks in 
data parallelism by
assigning uneven batch sizes to GPUs based on their relative compute speed. However, a GPU
with a high compute-to-memory ratio may not have enough memory to fully utilize its compute without
running out of memory.

In pipeline parallelism \cite{narayanan2019pipedream, park2020hetpipe}, balancing compute latency across stages is crucial, as the slowest stage bottlenecks the pipeline.
In homogeneous clusters, dividing the layers evenly across stages is effective
since transformer layers are typically identical \cite{narayanan2021memory}. In heterogeneous clusters, 
layers can be partitioned based 
on the relative compute speed of the GPUs. 

  However, achieving an efficient partition that balances compute may not be possible, 
as the fastest GPUs may lack sufficient memory to handle the layers required to maximize 
their compute potential, while slower GPUs might fully utilize their compute capacity but 
leave a significant portion of their memory underutilized.
HAP \cite{hap} distributes workloads unevenly in data and tensor parallelism to align with GPU compute capacities, though it still assumes faster GPUs have more memory.
Additionally, tensor parallelism requires high-bandwidth GPU interconnects for efficiency, which are unlikely to be available in heterogeneous clusters with lower-end GPUs.
Metis \cite{metis} and FlashFlex \cite{flashflex} integrate heterogeneous data, pipeline, and tensor (3D) parallelism, offering greater flexibility for heterogeneous training configurations but inherit the limitations of each parallelism type.

In existing data and model parallelism approaches, compute and memory allocation are tightly coupled, which becomes problematic in heterogeneous clusters since a GPU's memory capacity does not always match its compute speed (Fig. \ref{fig:gpu_latency_vs_memory}). This mismatch often prevents effective compute balancing, due to
memory limitations. \system solves these problems by independently 
balancing compute and memory during training in heterogeneous GPU clusters. 
\system targets the training of medium sized models, such as Llama and Phi, which offer competitive performance comparable to larger models \cite{abdin2024phi, size-importance, zhang2024tinyllama}. These models are feasible to train on moderately sized heterogeneous clusters, making them attractive options for organizations that seek high-performance models without large, high-end homogeneous GPU clusters.

\section{\system Design}
\system is designed to maximize training throughput by effectively balancing computational and memory loads
across heterogeneous GPUs, ensuring full utilization of the aggregate resources available in the cluster.

\begin{figure}[htbp]
  \centering
  \includegraphics[width=\columnwidth]{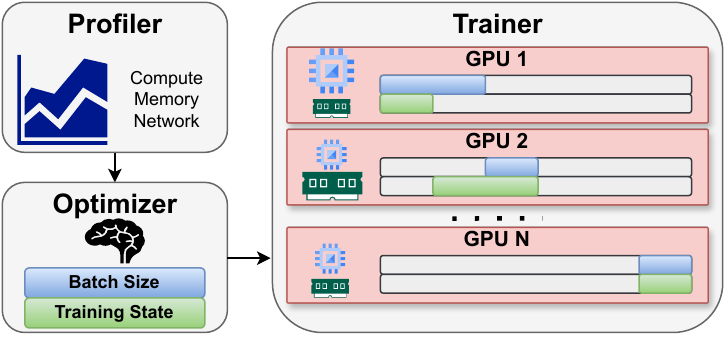}
  \caption{Architecture of \systemns.}
  \label{fig:architecture}
\end{figure}

\system is built on top of FSDP \cite{fsdp}, which divides the training state and computation evenly
across each GPU. To balance compute, \system assigns a batch size to each GPU proportional to its compute speed. 
To balance memory utilization, \system partitions the training state and decides on configurations for
gradient accumulation, activation checkpointing, and activation offloading 
according to the relative memory capacities of each GPU.
Given a model and target cluster, \system profiles the model to build performance models predicting computation time, memory usage, and communication time across configurations. The optimizer then leverages these models to configure batch size, training state shard, and gradient accumulation for each GPU to maximize training throughput. Figure \ref{fig:architecture} illustrates \systemns's architecture.

\subsection{Division of Compute and Training State}
A key feature of \system is its ability to decouple the distribution of compute and memory loads across GPUs, essential for optimizing performance in heterogeneous clusters where GPU memory capacity does not necessarily scale with compute power. \system efficiently allocates compute and training state across GPUs, leveraging the combined compute and memory resources of the cluster.
We describe the mechanisms \system
uses for this division next.

\textbf{\textit{Compute Partitioning.~}} Given a global batch size $B$, \system partitions the workload across GPUs by assigning each GPU $i$ a local batch size $b_i$ such that $\sum_i b_i = B$. To minimize iteration times, \system balances $b_i$ to reduce the maximum runtime on any GPU. To maintain equivalency with standard training, each GPU adjusts its local gradient by $N \cdot b_i / B$, resulting in a final gradient of:
\begin{equation}
  \nabla = \frac{1}{N}\sum_{i=1}^{N}(\frac{N\cdot b_i}{B})\frac{1}{b_i}\sum_{j=1}^{b_i}\nabla_{ij} = \frac{1}{B}\sum_{i=1}^{N}\sum_{j=1}^{b_i}\nabla_{ij}~~,
\end{equation}
where $\nabla_{ij}$ is the gradient on the $j$-th data input of GPU $i$.

\begin{figure*}[]
  \centering
  \includegraphics[width=\textwidth]{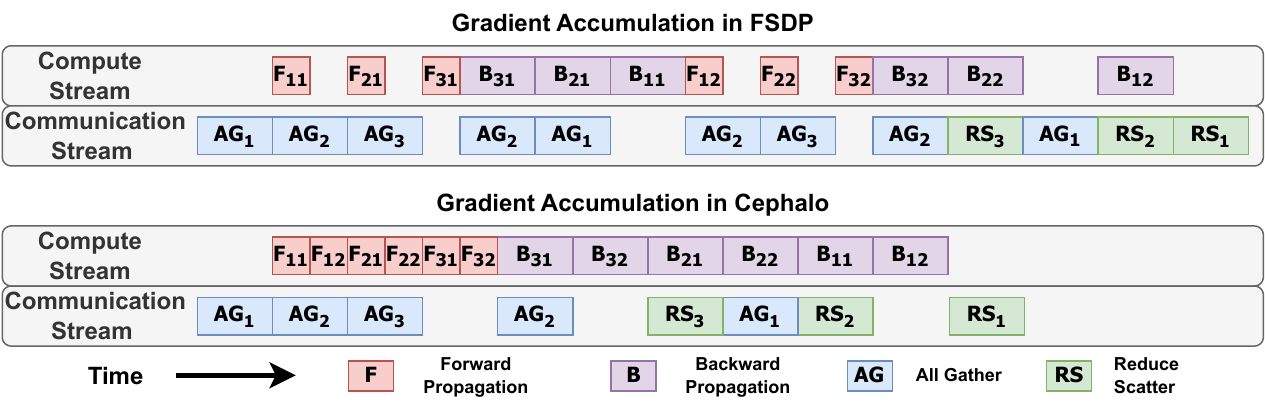}
  \caption{Gradient accumulation in FSDP (top) vs \system (bottom). The diagram illustrates gradient 
  accumulation over $2$ microbatches on a model consisting of $3$ FSDP units. $F_{ij}$ and $B_{ij}$
  are the forwards and backwards passes of the $i$th FSDP unit on the $j$th microbatch. $AG_i$ and $RS_i$
  are the \textit{AllGather} and \textit{ReduceScatter} collectives for the $i$th FSDP unit.}
  \label{fig:ga}
\end{figure*}

\textbf{\textit{Training State Partitioning.~}} The training state includes model parameters, gradients, and optimizer states, which consume significant memory during training. In FSDP, this state is evenly divided across GPUs, with each of the $N$ GPUs managing $1/N$ of the parameters and corresponding optimizer state throughout training. Model parameters are grouped into FSDP units, where compute and communication are managed collectively. During forward and backward passes, an \textit{AllGather} collective operation assembles the full parameter set on each GPU, and afterward, parameters are resharded to ensure only one unit is materialized in memory at any time. After each unit's backward pass, a \textit{ReduceScatter} collective averages gradients and sends them to the GPU responsible for those parameters.
Instead of a fixed partition, \system assigns each GPU $i$ a training state ratio $r_i$ such that $\sum r_i = 1$, allowing fine-grained memory control for each GPU independent of compute distribution.

\subsection{Managing Memory for Compute}
Beyond storing training state, significant GPU memory is needed for computation and storing intermediate activations. 
We employ gradient accumulation \cite{narayanan2019pipedream,lamy2021layered} to enable training with larger effective batch sizes while reducing memory usage. Instead of computing gradients for the full batch size $b$ at once, we split $b$ into smaller microbatches of size $m$ and accumulate gradients over $\ell$ microbatches, where $b = \ell \cdot m$. This approach allows each GPU to process an effective batch size of $b$ while reducing memory demands by managing smaller microbatches. In \system we develop an optimized implementation of gradient accumulation for FSDP, and configure it to control the amount of memory used for computation. 

\textbf{\textit{Layered Gradient Accumulation.~}} Traditional gradient accumulation in FSDP performs the full forward and backward pass for each microbatch sequentially. This necessitates $
\ell$ times more \textit{AllGather} collectives due to the need 
to gather sharded parameters for each microbatch. To mitigate this overhead, we implement \textit{layered} gradient accumulation \cite{lamy2021layered}, which processes \textit{all} microbatches for a given layer before moving to the next. Sequentially processing all microbatches
allows us to gather layer parameters only once for all microbatches per pass.

Figure \ref{fig:ga} illustrates the difference between gradient accumulation in FSDP and
\systemns. Our implementation calls \textit{AllGather} to prefetch the next FSDP unit 
while the current one is executing. This communication is overlapped with all executing microbatches of the 
current FSDP unit, effectively hiding the communication overhead even when networking is slow relative to compute.
Gradient accumulation can add minor runtime overhead as smaller microbatches may not fully utilize GPU cores, introducing a tradeoff between memory savings and compute efficiency. Unlike previous systems, \system automatically optimizes gradient accumulation with compute and training state partitioning (described in Section \ref{sec:optimizer}), balancing this tradeoff effectively.

\textbf{\textit{Activation Checkpointing and Offloading.~}} 
While layered gradient accumulation reduces communication overhead, it introduces significant memory overhead compared to traditional gradient accumulation. This is because activations must be stored for all microbatches of a layer until the backward pass, whereas traditional gradient accumulation only maintains activations for a single microbatch. For some models, this additional activation storage can exceed the memory savings gained from smaller batch sizes.

\system addresses memory overhead in layered gradient accumulation with a combination of activation checkpointing and offloading. Activation checkpointing saves activations only at layer boundaries during the forward pass \cite{pipedream,shoeybi2019megatron}, allowing intermediate activations to be recomputed in the backward pass, which significantly reduces memory usage. However, even storing boundary activations adds overhead. To mitigate this, \system uses activation offloading to move boundary activations to CPU memory until needed in the backward pass.
PyTorch's default activation offloading was too slow due to synchronous CPU-GPU transfer, which blocked GPU computation. Consequently, we developed an optimized asynchronous offloading method that transfers activations between GPU and CPU while computations continue, eliminating memory overhead in layered gradient accumulation with minimal latency. Section \ref{sec:trainer} details this offloading strategy and additional optimizations that were necessary to run layered accumulation efficiently with FSDP.

\subsection{Performance Modeling}
The profiler runs training iterations on small batch sizes in the target cluster to build predictive models for compute latency and memory usage based on batch size. We use linear models, as they are simple, require minimal profiling to fit, and accurately predict both metrics. We profile communication latency for collectives with an evenly sharded training state and apply a conservative model to adjust for latency when the state is unevenly sharded. The optimizer then uses these models to find a configuration that maximizes throughput while respecting each GPU’s memory capacity.

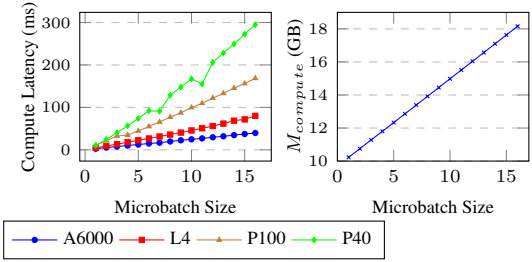
\begin{figure}[htbp]
  \centering
  \begin{tikzpicture}
    \begin{groupplot}[
      group style={
          group size=2 by 1,
          horizontal sep=25pt, 
      },
      ylabel style={yshift=-2mm}, 
      width=0.5\linewidth, 
  ]

    \nextgroupplot[
        xlabel={\scriptsize Microbatch Size},
        ylabel={\scriptsize Compute Latency (ms)},
         y tick label style={font=\tiny},
          x tick label style={font=\scriptsize},
        legend style={
            font=\scriptsize, 
            at={(-.4,-0.4)}, 
            anchor=north west,
            legend columns=-1, 
        },
        ymajorgrids=true,
        grid style=dashed,
        mark size=1pt,
    ]
    
    \addplot[blue,mark=*,] coordinates {
      (1,2.6684666666666677)
      (2,5.307666666666663)
      (3,7.727400000000002)
      (4,10.110333333333328)
      (5,12.506666666666668)
      (6,14.848666666666665)
      (7,16.781200000000016)
      (8,19.654733333333336)
      (9,22.3898)
      (10,24.68126666666665)
      (11,26.936599999999988)
      (12,29.503000000000025)
      (13,31.813933333333342)
      (14,34.62546666666667)
      (15,37.257933333333334)
      (16,39.725666666666626)
  };
  \addlegendentry{A6000}
  
  \addplot[red,mark=square*] coordinates {
      (1,4.4866)
      (2,9.13826666666667)
      (3,13.295866666666681)
      (4,17.980066666666655)
      (5,22.497866666666653)
      (6,26.743466666666656)
      (7,31.390799999999963)
      (8,35.88026666666667)
      (9,40.6432)
      (10,45.64086666666671)
      (11,50.70306666666665)
      (12,55.99819999999998)
      (13,61.76919999999999)
      (14,68.67740000000003)
      (15,71.69846666666666)
      (16,80.09113333333335)
  };
  \addlegendentry{L4}
  
  \addplot[brown,mark=triangle*] coordinates {
      (1,10.560066666666662)
      (2,21.584733333333343)
      (3,32.63373333333337)
      (4,34.79260000000002)
      (5,44.405133333333325)
      (6,54.83866666666661)
      (7,65.63440000000001)
      (8,76.89766666666664)
      (9,87.26673333333329)
      (10,99.50853333333336)
      (11,109.60400000000006)
      (12,121.79140000000001)
      (13,132.94360000000003)
      (14,145.4232666666665)
      (15,156.2774)
      (16,168.5038)
  };
  \addlegendentry{P100}
  
  \addplot[green,mark=diamond*] coordinates {
      (1,10.073466666666672)
      (2,24.553266666666655)
      (3,40.58693333333333)
      (4,56.66066666666665)
      (5,73.8723333333333)
      (6,92.03166666666667)
      (7,90.7959333333333)
      (8,128.84013333333345)
      (9,147.78813333333318)
      (10,167.31979999999993)
      (11,154.88966666666659)
      (12,206.27320000000003)
      (13,228.06340000000012)
      (14,249.4232666666667)
      (15,272.5648)
      (16,294.6486000000001)
  };
  \addlegendentry{P40}

       \nextgroupplot[
        xlabel={\scriptsize  Microbatch Size},
        ylabel={\scriptsize $M_{compute}$ (GB)},
        xmin=0, xmax=17,
        ymin=10, ymax=19,
         y tick label style={font=\tiny},
          x tick label style={font=\scriptsize},
        ymajorgrids=true,
        grid style=dashed,
        mark size=1pt,
    ]
    \addplot[color=blue, mark=x,]
    coordinates {
    (1,10.23)(2,10.75)(3,11.28)(4,11.8)(5,12.33)(6,12.86)(7,13.39)(8,13.92)(9,14.45)(10,14.98)(11,15.51)(12,16.04)(13,16.57)(14,17.1)(15,17.63)(16,18.16)
    };

  \end{groupplot}
    \end{tikzpicture}
  \caption{Training latency and memory allocated for compute as the microbatch size increases for Bert-Large.}
  \label{fig:memory_vs_batch}
\end{figure}

\textbf{\textit{Compute Latency Model.~}}
In the left plot of Figure \ref{fig:memory_vs_batch}, we profile the compute latency of a single transformer layer
as the batch size increases. For small batch sizes, the latency increases sublinearly as the batch size is not
large enough to fully utilize the compute on the GPU. As the GPU compute is saturated for larger batch sizes, there is a strong linear relationship.
We model latency by using the profiled data for smaller batches to capture non-linearities, then extrapolate linearly for larger batches. Profiling a single layer reduces time and resources, and since transformer layers are typically identical, we can extrapolate the entire model’s latency from a single layer's profile.

Let $T_f(m)$ and $T_b(m)$ be the latency models for
forwards and backwards compute as a function of the microbatch size $m$.
We linearly scale the latency of a single microbatch by the number of microbatches $\ell$ 
to derive the total forwards $T_f(m,\ell)$ and backwards $T_b(m,\ell)$ compute latencies.

\textbf{\textit{Memory Utilization Model.~}}
\label{sec:memory_model}
During training, GPU memory utilization includes memory for the training state, $M_{state}$, and computation, $M_{compute}$, resulting in a total memory usage of $M=M_{state}+M_{compute}$.
$M_{state}$ is derived from the parameters in a GPU's model shard $|P|$. We assume standard full-precision training with the Adam optimizer, where each parameter requires $4$ bytes for the parameter, $4$ bytes for its gradient, and $8$ bytes for the first and second gradient moments. Thus, the total memory needed for the training state is $M_{state} = 16 \cdot |P|$.
$M_{compute}$ encompasses memory for executing GPU kernels, storing activations, and other framework state. In the right plot of Figure \ref{fig:memory_vs_batch}, we plot $M_{compute}$ against batch size by subtracting $M_{state}$ from the total memory usage, showing a strong linear relationship. We profile $M_{compute}$ for small batch sizes to fit a linear model based on microbatch size. The linear increase is due to the need to run kernels and store activations for larger batch sizes. Notably, $M_{compute}$ is unaffected by the number of microbatches, as activations are checkpointed and offloaded after computation in \systemns.

\textbf{\textit{Communication Latency.~}}
\label{sec:comm_model}
FSDP uses NCCL \cite{NVIDIA_NCCL_2024} for inter-GPU communication, using \textit{AllGather} to collect parameters and \textit{ReduceScatter} to average gradients. 
With even training state sharding, inputs to NCCL collectives are equal in size; however, uneven sharding introduces variable input sizes. \system employs generalized collective implementations which handle uneven inputs but incur overhead from extra GPU memory copies \cite{fsdp}.
In practice, the overhead from uneven input sizes remained within $15\%$ of even sharding, shown in Supplementary Material Section C. Therefore, we profile collective latency with even inputs and assume a conservative $15\%$ overhead for uneven sharding. Since transformer layers are identical, profiling is needed only for a single layer.

\subsection{Optimizer}
\label{sec:optimizer}
Given a model, cluster of $N$ GPU machines, and a target batch size $B$ to train with, 
the optimizer decides how to divide the computation, training state,
and configure gradient accumulation to maximize training throughput. 
Next, we describe how the optimizer formulates this as an optimization problem
and solves it with dynamic programming.

\textbf{\textit{Optimization Formulation.~}}
We maximize training throughput by minimizing the latency for one iteration of training.
Under the typical assumption that transformer layers are identical, this problem is equivalent to
minimizing the forwards and backwards pass for a single transformer layer. 
We wrap each transformer layer as an FSDP unit~\cite{pytorch_trillion_model}, which efficiently overlaps communication during the forwards and backwards pass. The forwards pass runs in 
\begin{equation}
  \label{eq:forward_latency}
  T_f = \max(\max_{i} (T_f^{g_i}(m_i, \ell_i)), AG),
\end{equation}

\begin{table}[h!]
\centering
\resizebox{\columnwidth}{!}{%
\begin{tabular}{|c|l|}
\hline
\textbf{Symbol} & \textbf{Description} \\ \hline
$N, B$           & Number of layers and GPUs \\ \hline
$m_i, \ell_i$, $g_i$   & Microbatch size and number of microbatches for $i$th GPU $g_i$ \\ \hline
$M(m)$          & Compute memory for microbatch size $m$ \\ \hline
$M_{cap}^{g_i}$ & Memory capacity of $g_i$ \\ \hline
$T^{g_i}_f(m, \ell)$ & Forwards latency of $g_i$ for $\ell$ microbatches of size $m$ \\ \hline
$T_b^{g_i}(m, \ell)$ & Backwards latency of $g_i$ for $\ell$ microbatches of size $m$ \\ \hline
$AG$, $RG$           & \textit{AllGather} and \textit{ReduceScatter} latency \\ \hline
$M_{state}^{es}$ & Memory required to store an even training state share \\ \hline
\end{tabular}%
}
\caption{Notation and Definitions}
\label{table:notation}
\end{table}
where variables are defined in Table \ref{table:notation}.
The forwards pass waits on the slowest GPU to finish its computation, as well as the \textit{AllGather} that is running concurrently to fetch the next FSDP unit. Similarly, the backwards pass will take
\begin{equation}
  \label{eq:backward_latency}
  T_b = \max(\max_{i} (T_b^{g_i}(m_i, \ell_i)), RS + AG),
\end{equation}
where a \textit{ReduceScatter} is required to average the gradient. The training state must be unevenly sharded if, for any GPU, its combined compute memory and the evenly distributed training state memory exceeds its memory capacity.
Then, the goal is to minimize the layer latency $T_f + T_b$ subject to the constraints: (I) Batch size: $ B = \sum_i b_i=m_i\cdot \ell_i, \ell_i \in \mathbb{Z}_{>0} $ (II) Individual memory: $M(m_i) \le M_{cap}^{g_i}, \forall i$ (III) Aggregate memory: $M_{state} + \sum_i M(m_i) \le \sum_i M_{cap}^{g_i}$.\\
The second constraint specifies that the memory used for compute cannot exceed the memory capacity of the GPU.
The last constraint specifies that the aggregate GPU memory in the cluster is at least as much as the sum of the 
memory required to store the complete training state and perform computation on each GPU. Under these conditions, \system is able to train the model without running out of memory.

\textbf{\textit{Dynamic Programming Solution.~}}
We solve the optimization problem using 
dynamic programming. 
Let $D(i,j,k)$ be the minimum achievable runtime for the first $i$ GPUs to process a total batch size of $j$
and total microbatch size of $k$.  That is, the sum of the batch sizes on the first $i$ GPUs is $j$, and the sum of
their microbatch sizes is $k$.
Suppose that the optimal solution assigns $\ell$ microbatches of size $m$ (batch size of $\ell\cdot m$) to the $i$th GPU. 
Then the optimal solution can 
be constructed by combining this assignment with the solution to $D(i-1,j-\ell\cdot m,k-m)$. Thus, by this optimal
subproblem property, we can compute $D(i, j, k)$ as 
\begin{equation}
  D(i, j, k) = \min_{m,\ell} \max(D(i-1, j-\ell\cdot m, k-m), T_{i,\ell, m})~,
\end{equation}
where $\ell\cdot m \le j, m \le k, M(m) \le M_{cap}^{g_i}$
and $T_{i,\ell, m}$ is the runtime of forwards and backwards for $\ell$ microbatches of size $m$ on the $i$th GPU using Eqs. \ref{eq:forward_latency} and \ref{eq:backward_latency}.

From our memory model, we can compute the aggregate memory utilization using
the sum of the microbatch sizes, $k$. Hence, the last dimension in the recurrence represents the aggregate memory utilization.
This dimension is needed in the recurrence to ensure constraint (III) is satisfied. 
The minimum latency is $\min_{k} D(N, B, k)$ over all $k$ meeting the memory constraint. We then backtrack to find the batch and microbatch sizes that achieve this throughput. Pseudocode is in Supplementary Material Section A.1, with experiments validating the model's accuracy in Section A.3.

\textbf{\textit{Training State Partition.~}}
After determining the compute partitioning, the optimizer allocates training state to minimize the maximum memory utilization across GPUs, balancing each GPU's memory consumption relative to its capacity. This prevents out-of-memory issues and reduces memory allocation overheads when memory utilization approaches capacity. This allocation is computed using a greedy algorithm, assigning training state iteratively to the GPU with the lowest memory utilization until fully distributed.

\textbf{\textit{Complexity Analysis.~}}
The optimizer runtime is dominated by the dynamic programming algorithm which runs in $O(N \cdot B^3 \cdot \log B)$, where $N$ is the GPU count and $B$ the global batch size. This arises from $O(N \cdot B^2)$ states, each requiring $O(B \cdot \log B)$ to compute. The greedy algorithm for training state partitioning runs in $O(N^2)$.

\section{Implementation}
\system is implemented on top of FSDP in 
PyTorch and consists of a profiler, optimizer and model trainer (Fig. \ref{fig:architecture}). This section details \systemns's implementation and optimizations.

\subsection{Profiler}
\label{profiler}
The profiler performs lightweight profiling to model compute latency, memory usage, and communication latency. It profiles a few training iterations for each batch size from $1$ to $B$, fitting linear models for compute latency and memory usage. In practice, $B=8$ suffices for accuracy. The profiler also measures \textit{AllGather} and \textit{ReduceScatter} latencies.

\subsection{Optimizer}
The optimizer uses models built by the profiler to configure \system for maximum training throughput (Section \ref{sec:optimizer}). It determines each GPU's microbatch size, number of microbatches, and assigned portion of the global batch size and training state. To avoid memory allocation bottlenecks as usage nears capacity, the optimizer caps GPU memory usage at 80\%. It runs within 20 minutes for all workloads, which is negligible relative to the GPU-years required to train these models \cite{touvron2023llama}. Supplementary Material Section A.2 details the optimization time breakdown.
\vspace{-0.5em}

\subsection{Trainer}
\label{sec:trainer}
\textbf{\textit{Compute and Training State Division.}}~
The \textit{trainer} trains the model using the batch size and training state assignments set by the optimizer. Each process’s data loader is configured to load its assigned batch size. \systemns's training logic is compatible with any sequential model defined in PyTorch. This applies to transformer models, which are typically structured as a sequence of identical layers.

\textbf{\textit{Uneven Parameter Sharding.}}~
Implementing uneven parameter sharding required modifying FSDP’s \textit{shard} and \textit{unshard} operations to follow the training state divisions configured by the optimizer. The gradient synchronization logic in the backward pass was also updated to average gradients according to this training state division.

When the training state is unevenly sharded, \system uses 
generalized \textit{AllGather} and \textit{ReduceScatter} 
implementations to handle uneven input sizes. We observed uneven sharding incurs up to a 15\% 
runtime overhead, but does not have a strong correlation with the skew in shard sizes. Therefore, we apply a greedy strategy to minimize uneven sharding across FSDP units. For instance, if two identical FSDP units are split across two GPUs in a 3:1 ratio, we would shard one unit evenly (1:1) and the other as 1:0, incurring uneven sharding overhead for only one unit.

\textbf{\textit{Layered Gradient Accumulation.}}~
The trainer implements a training loop for layered gradient accumulation, splitting each batch into microbatches and processing them one at a time through each FSDP unit. For both forward and backward passes, it runs all microbatches on one FSDP unit before moving to the next. This order differs from FSDP’s assumed layer-wise sequential execution, which it uses to overlap communication with computation. Consequently, several changes were needed in FSDP to avoid unnecessary communication and support communication-computation overlap with this new order of execution.

FSDP reshards parameters after each forward pass, assuming the next unit runs next and the current is no longer needed. However, in gradient accumulation, the same unit runs all microbatches before moving to the next. We modified FSDP to reshard parameters only after all microbatches are processed, avoiding unnecessary \textit{AllGather} operations. To maintain communication-computation overlap, we updated the prefetching logic to align with the layered gradient accumulation order and scheduled unsharding logic on a separate GPU stream to avoid blocking the next microbatch’s backward computation. Finally, we adjusted post-backward logic to accumulate gradients across microbatches and reset the execution state only after all microbatches are processed.

We observed severe memory fragmentation from PyTorch scheduling multiple microbatches simultaneously, leading to out-of-memory errors even below 50\% memory usage. We avoid this fragmentation by synchronizing the GPU's compute stream to process one microbatch at a time.

Lastly, layered gradient accumulation raises memory overhead by requiring activations to be held until all microbatches are processed. We avoid this overhead by checkpointing and asynchronously offloading activations and gradients to CPU memory when unused. Supplementary Material (Section B) provides more details. In Section \ref{sec:ablation_study}, we show that our optimized layered gradient accumulation with checkpointing and offloading is essential for performance.

\section{Performance Evaluation}
We evaluate the performance of \system compared to state-of-the-art 
training methods on $9$ popular transformer models across $2$ heterogeneous GPU clusters.
End-to-end results are presented in Section \ref{sec:end_to_end}, and larger-scale experiments in Section \ref{sec:larger_cluster}. Sections \ref{sec:ablation_study} and \ref{sec:grad_accum} analyze how \systemns's design components impact performance. 
Section \ref{sec:opt_config} presents training configurations generated by \systemns.

\subsection{Experimental Setup}
\label{sec:experimental_setup}
We evaluate popular transformer models used for text classification (TC), text generation (TG), and image 
classification (IC) following the training setup from \cite{pytorch_trillion_model}. Activations are checkpointed after each transformer layer and models are trained in full precision with the Adam optimizer, using a sequence length of 512 for language modeling.
Table \ref{fig:one_col_model_specs} provides further details on the models.

\textbf{\textit{Clusters.}}~
We evaluated \system on environments representative of typical heterogeneous GPU clusters used by ML practitioners. Cluster A was assembled with four types of GPUs acquired over several years. Cluster B is a mix of higher- and lower-end GPU VMs on AWS, selected to reflect the typical quantities available for reservation.

\begin{itemize}
  \item Cluster A: $2$ machines ($8$ GPUs), connected via a 50 Gbps link. One contains 2$\times$L4, 1$\times$A6000, 
  and 1$\times$P40; the other contains 2$\times$P40 and 2$\times$P100. 
  \item Cluster B: $8$ VMs ($64$ GPUs), equipped with 100 Gbps bandwidth. 2$\times$g5.48xlarge (8$\times$A10G), 
  2$\times$p3.16xlarge (8$\times$V100-16GB), and 4$\times$g4dn.metal (8$\times$T4) VMs.
\end{itemize}
A summary of GPU specifications appear in Table \ref{tab:gpu_specs}.

\textbf{\textit{Baselines.}}~
We compare against representative state-of-the-art techniques for training on heterogeneous GPU clusters:
\begin{itemize}
    \item Megatron-Het \cite{megatron-lm}: Employs pipeline parallelism across nodes and data/tensor parallelism within nodes. We adapted it for heterogeneous training by partitioning the model proportionally to each node's compute capacity.
    \item  FlashFlex \cite{flashflex}: Combines ZeRO-2 data \cite{zero3} (optimizer state and gradient sharding), tensor, and pipeline parallelism. An optimizer balances memory and compute across GPUs.
\end{itemize}
Whale \cite{whale}, HAP \cite{hap}, and baseline FSDP, which ran out of memory on most workloads, are compared in Supplementary Material Section D.

\begin{table}[ht]
  \centering
  \caption{Model Statistics}
  \vspace{0.2em}
  \renewcommand{\arraystretch}{1.1}
  \setlength{\extrarowheight}{2pt}
  \footnotesize 
  \resizebox{\linewidth}{!}{%
   \begin{tabular}{|c|l|c|c|c|c|}
    \hline
    \textbf{Task} & \textbf{Model} & \textbf{Layers} & \textbf{Embd. Size} & \textbf{Attn. Heads} & \textbf{Parameters} \\ \hline
    IC & ViT-G \cite{zhai2022scaling} & 48 & 1664 & 16 & 1.8B \\
    IC & ViT-e \cite{chen2022pali} & 56 & 1792 & 16 & 3.9B \\ \hline
    TC & BERT-Large \cite{devlin2018bert} & 24 & 1024 & 16 & 0.4B \\
    TC & BERT-XLarge \cite{devlin2018bert} & 36 & 1536 & 24 & 1.2B \\ 
    \hline
    TG & GPT 2.7B \cite{gpt} & 32 & 2560 & 80 & 2.7B \\
    TG & GPT 6.7B \cite{gpt} & 32 & 4096 & 128 & 6.7B \\
    TG & Tiny Llama \cite{zhang2024tinyllama}  & 22 & 2048 & 32 & 1.1B \\
    TG & Llama 3B \cite{openlm2023openllama} & 26 & 3200 & 32 & 3.5B \\
    TG & Llama 7B \cite{touvron2023llama} & 32 & 4096 & 32 & 6.7B \\ \hline
    \end{tabular}}
  \label{fig:one_col_model_specs}
\end{table}
\vspace{-0.5cm}
\begin{table}[ht]
\centering
\caption{GPU Specifications}
\vspace{0.2em}
\renewcommand{\arraystretch}{1.05}
\setlength{\extrarowheight}{0.5pt}
\resizebox{0.8\linewidth}{!}{%
\begin{tabular}{|c|l|l|c|c|}
\hline
\textbf{Cluster} & \textbf{GPU} & \textbf{Generation} & \textbf{Memory} & \textbf{TFlops (FP32)} \\ \hline
\multirow{4}{*}{A} & P40          & Pascal              & 24 GB     & 11.8                     \\ 
                   & P100         & Pascal              & 12 GB      & 9.3                    \\ 
                   & A6000        & Ampere              & 48 GB     & 38.7                   \\ 
                   & L4           & Ada                 & 24 GB     & 30.3                   \\ \hline
\multirow{3}{*}{B} & V100         & Volta               & 16 GB   & 14.1                    \\ 
                   & T4           & Turing              & 15 GB     & 8.1                    \\ 
                   & A10G         & Ampere              & 24 GB     & 31.2                   \\ \hline
\end{tabular}}
\label{tab:gpu_specs}
\end{table}

\begin{table*}[ht]
\centering
\footnotesize
\caption{Throughput comparison of different models and batch
 sizes on 8-GPU Cluster A. \textit{OOM} denotes Out-of-Memory.}
\label{table:model_strategy_performance}
\renewcommand{\arraystretch}{0.8}
\resizebox{\textwidth}{!}{
\begin{tabular}{@{}lrrrrrrrrrrrrrrrr@{}}
\toprule
\textbf{System} & \multicolumn{2}{c}{\textbf{ViT-G}} & \multicolumn{2}{c}{\textbf{ViT-e}} & \multicolumn{2}{c}{\textbf{Bert-Large}} & \multicolumn{2}{c}{\textbf{Bert-XLarge}} & \multicolumn{2}{c}{\textbf{GPT 1.3B}} & \multicolumn{2}{c}{\textbf{GPT 2.7B}} & \multicolumn{2}{c}{\textbf{Tiny Llama}} & \multicolumn{2}{c}{\textbf{Llama 3B}} \\ 
\cmidrule(lr){2-3} \cmidrule(lr){4-5} \cmidrule(lr){6-7} \cmidrule(lr){8-9} \cmidrule(lr){10-11} \cmidrule(lr){12-13} \cmidrule(lr){14-15} \cmidrule(lr){16-17}
& \textbf{128} & \textbf{256} & \textbf{128} & \textbf{256} & \textbf{128} & \textbf{256} & \textbf{128} & \textbf{256} & \textbf{128} & \textbf{256} & \textbf{128} & \textbf{256} & \textbf{128} & \textbf{256} & \textbf{128} & \textbf{256} \\ 
\midrule
Megatron-Het & 3.41 & 0.79 & \textit{OOM} & \textit{OOM} & 19.77 & 20.57 & 6.40 & 6.80 & 4.18 & 4.35 & 1.82 & 1.82 & 7.93 & 8.63 & \textit{OOM} & \textit{OOM} \\ 
FlashFlex & 2.88 & 2.97 & 1.38 & 1.4 & 25.64 & 28.90 & 8.63 & 9.06 & 5.81 & 5.83 & 2.79 & 2.83 & 8.67 & 8.75 & 1.91 & 1.83 \\ 
\system & \textbf{6.38} & \textbf{6.41} & \textbf{3.02} & \textbf{3.23} & \textbf{33.56} & \textbf{33.69} & \textbf{11.47} & \textbf{11.72} & \textbf{6.83} & \textbf{7.09} & \textbf{4.57} & \textbf{4.67} & \textbf{12.58} & \textbf{12.91} & \textbf{4.51} & \textbf{4.85} \\
\bottomrule
\end{tabular}
}
\end{table*}

\subsection{Training Throughput}
\label{sec:end_to_end}
We evaluated \systemns's end-to-end training throughput against baselines, measuring throughput as samples processed per second (images for image classification models, sequences for language models). Experiments on Cluster A included models up to 3.9 billion parameters with global batch sizes of 128 and 256. Cluster A is highly heterogeneous, with four GPU types varying substantially in compute and memory. Baselines do not auto-configure pipeline parallelism, 
so we tested various microbatch sizes (powers of 2), with the best results reported in Table \ref{table:model_strategy_performance}. \system consistently achieved significantly higher throughput without out-of-memory (\textit{OOM}) errors across all models and batch sizes. 

\textbf{\textit{Comparison to Megatron-Het.~}}
Megatron uses four pipelines of two GPUs each across the two nodes. However, each pipeline must be partitioned identically, despite the mixed GPU types on each node. This results in different GPUs being assigned the same stage across pipelines, causing compute bottlenecks due to the slower P40 GPUs, which underutilizes faster L4 and A6000 GPUs, reducing throughput. For larger models (GPT 2.7B and Llama 3B), Megatron applies tensor parallelism within each node, further decreasing throughput due to high communication overhead. Megatron is optimized for clusters with fast interconnects like NVSwitch, which Cluster A and most AWS VMs do not have (except mostly unavailable A100 and H100 VMs).

\textbf{\textit{Comparison to FlashFlex.~}}
Like \systemns, FlashFlex trains larger batch sizes with a reduced memory footprint by using smaller microbatches and gradient accumulation. However, smaller microbatches may not fully utilize GPU compute, and frequent gradient accumulation reduces pipeline parallelism efficiency. \system automatically optimizes the microbatch size and gradient accumulation configuration, whereas FlashFlex requires manual tuning. Additionally, \systemns’s layered gradient accumulation implementation does not incur extra communication overhead. FlashFlex, like Megatron-Het, relies on communication-heavy tensor parallelism for larger models. These factors enable \system to achieve significantly higher throughput across all configurations.
\begin{table}[]
\centering
\small
\caption{Throughput comparison on 64-GPU Cluster B.}
\label{tab:model_strategy_batch_performance}
\renewcommand{\arraystretch}{0.8}
\resizebox{\linewidth}{!}{%
\begin{tabular}{lrrrrrr}
\toprule
\textbf{System} & \multicolumn{2}{c}{\textbf{ViT-e}} & \multicolumn{2}{c}{\textbf{GPT 6.7B}} & \multicolumn{2}{c}{\textbf{Llama 7B}}  \\
\cmidrule(lr){2-3} \cmidrule(lr){4-5} \cmidrule(lr){6-7}
         & \textbf{512} & \textbf{1024} & \textbf{512} & \textbf{1024} & \textbf{512} & \textbf{1024} \\
\midrule
Megatron-Het     & 12.06 & 12.12 & 3.59 & 1.71 & 5.53 & 1.65 \\
FlashFlex        & 12.84 & 13.37 & 4.78 & 4.99 & 5.42 & 5.47 \\
\system          & \textbf{20.37} & \textbf{26.08} & \textbf{11.62} & \textbf{17.04} & \textbf{13.12} & \textbf{17.74} \\
\bottomrule
\end{tabular}
}
\end{table}

\subsection{Larger Cluster Experiments}
\label{sec:larger_cluster}
We evaluated \systemns's scalability on the larger Cluster B featuring 64 GPUs (16 V100s, 16 A10Gs, and 32 T4s) using ViT-e, GPT-6.7B, and Llama-7B models with batch sizes of 512 and 1024. \system consistently delivered 2-10$\times$ higher throughput than other systems.

\begin{figure}
  \begin{tikzpicture}
    \begin{groupplot}[
      group style={
        group size=2 by 1,
        ylabels at=edge left,
        horizontal sep=1.2cm,
        vertical sep=0.6cm,
      },
      width=0.50\linewidth,
      ybar,
      ymin=0,
      ylabel style={yshift=-2mm}, 
      symbolic x coords={ViT-e, GPT 6.7B, Llama 7B},
      xtick=data,
      legend style={at={(0.6,-0.5)},anchor=north,legend columns=-1, font=\scriptsize},
      height=3cm,
      width=4.5cm,
      enlarge x limits=0.3,
      y tick label style={font=\tiny},
      x tick label style={font=\tiny, rotate=15, anchor=east, xshift=13, yshift=-3},
      ymajorgrids=true,
      grid style=dashed,
      scaled y ticks=false,
      x tick style={draw=none},
    ]
      \nextgroupplot[bar width=4pt,ylabel={\footnotesize TFLOPs}]
      \addplot+[fill=blue] coordinates {(ViT-e, 71.75) (GPT 6.7B, 133.76) (Llama 7B, 153.47)};
      \addplot+[fill=red] coordinates {(ViT-e, 104.86) (GPT 6.7B, 201.21) (Llama 7B, 224.32)};
      \addplot+[fill=yellow] coordinates {(ViT-e, 132.71) (GPT 6.7B, 253.26) (Llama 7B, 280.02)};
      \legend{A10G, A10G+V100, A10G+V100+T4} 
      \nextgroupplot[bar width=4pt,ylabel={\footnotesize TFLOPs}]
      \addplot+[fill=yellow] coordinates {(ViT-e, 132.71) (GPT 6.7B, 253.26) (Llama 7B, 280.02)};
      \addplot+[fill=brown] coordinates {(ViT-e, 135.91) (GPT 6.7B, 248.34) (Llama 7B, 276.36)};
      \legend{Heterog., Homog.} %

    \end{groupplot}
  \end{tikzpicture}
  \caption{Left: Throughput (TFLOPs) with different heterogeneous cluster configurations. 
  Right: Throughput (TFLOPs) on Cluster B vs. a homogeneous cluster of 32$\times$A10G GPUs.}
  \label{fig:scale_cluster_flipped}
  \vspace{-0.5cm}
\end{figure}
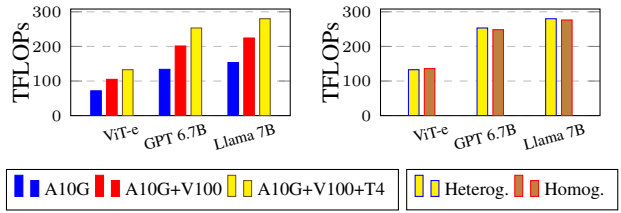

At a batch size of $512$, Megatron uses ZeRO-2 data parallelism within each node. Since it does not
shard the model parameters like \systemns, Megatron needs to configure pipeline parallelism with a smaller
microbatch size and a suboptimal model partitioning to avoid running out of memory. It is unable to fully utilize
compute on the V100 GPUs since it has similar memory to the T4 despite being significantly faster. At 
a batch size of $1024$, 
Megatron uses tensor parallelism to manage memory. However, this reduces throughput for GPT 6.7B and Llama 7B, as V100 GPUs' NVLink lacks all-to-all connectivity and is not fast enough to offset the communication overhead of tensor parallelism.

FlashFlex is able to more flexibly parallelize training, supporting a different degree of tensor
parallelism per pipeline stage and a different number of GPUs for each pipeline. This enables faster training at a batch size
of $1024$ when memory pressure is larger.  
However, it still relies on tensor parallelism (albeit less than Megatron) and partitions layers into pipeline stages according to memory, rather than compute, to avoid running out of memory. This partitioning assigns the T4s a similar workload as the V100s, despite being slower, resulting in a performance bottleneck.

In contrast, \system leverages FSDP to shard training state, reducing memory requirements and enabling training at a batch size of $1024$ without tensor parallelism. Additionally, independent partitioning of training state from compute allows \system to fully utilize each GPU by assigning batch sizes proportional to its compute capacity.

\textbf{\textit{Scaling Heterogeneous GPUs.~}}
In the left plot of Figure 
\ref{fig:scale_cluster_flipped}, we compare the training throughput (in TFLOPs) of \system as we scale from using
only the fastest A10G GPUs in Cluster B, to using the A10G and V100 GPUs, to finally using all GPUs.
The training throughput almost doubles when comparing only using A10G to utilizing all the heterogeneous
GPUs in the cluster. \system is able to achieve a significant improvement in 
training throughput by utilizing all of the (heterogeneous) GPUs available on the cluster.

\textbf{\textit{Comparison to Homogeneous Training.~}}
In the right plot of Figure 
\ref{fig:scale_cluster_flipped}, we compare \systemns’s training TFLOPs on Cluster B to a homogeneous cluster of 32$\times$A10Gs with similar peak TFLOPs (984 vs. 998). Despite Cluster B’s mix of lower-memory lower-compute GPUs, \system is able to achieve comparable TFLOPs to the homogeneous cluster, demonstrating effective utilization of heterogeneous GPUs. 

\subsection{Ablation Study}
\label{sec:ablation_study}
\begin{figure}
  \begin{tikzpicture}
  \begin{groupplot}[
      group style={
          group size=3 by 1,
          horizontal sep=14pt,
          ylabels at=edge left,
          xlabels at=edge bottom,
          vertical sep=10pt,
      },
      height=3.5cm,
      width=0.44\linewidth,
      title={Throughput vs. Global Batch Size},
      ylabel={\footnotesize Throughput},
      ylabel style={yshift=-2mm}, 
      yticklabel style={font=\tiny},
      xticklabel style={font=\tiny},
      xmin=0, xmax=260, ymin=0,
      xtick={0,64,128,192,256},
      legend pos=outer north east,
      legend style={
            font=\scriptsize, 
            at={(-1.1,-0.5)}, 
            anchor=north west,
            legend columns=-1, 
        },
      ymajorgrids=true,
      grid style=dashed,
  ]

    \nextgroupplot[title={\small ViT-e}]
  \addplot+[mark size=1.pt, mark=diamond*] coordinates {(16,1.46) (32,1.56) (48,1.53) (64,1.50) (80,1.47) (96,1.46)};
  \addplot+[mark size=1.pt, mark=triangle*] coordinates {(16,2.10) (32,2.97) (48,2.89) (64,3.07) (80,3.01) (96,2.91) (112,2.85) (128,2.86)};
  \addplot+[mark size=1.pt,mark=x] coordinates {(16,1.38) (32,1.84) (48,1.91) (64,1.95) (80,1.96) (96,1.98) 
    (112,2.00) (128,1.99) (144,1.99) (160,2.02) (176,2.00) 
    (192,2.01) (208,2.02) (224,2.01) (240,2.01) (256,2.02)};
  \addplot+[mark size=1.pt,mark=*] coordinates {(16,2.15) (32,2.82) (48,3.00) (64,3.08) (80,3.11) (96,3.16) 
    (112,3.14) (128,3.06) (144,3.16) (160,3.14) (176,3.27) 
    (192,3.22) (208,3.18) (224,3.18) (240,3.21) (256,3.23)};

  \node at (axis cs:96,1.46) [anchor=west] {\tiny\textcolor{blue}{\textit{OOM}}};
  \node at (axis cs:128,2.86) [anchor=west] {\tiny\textcolor{red}{\textit{OOM}}};
  
\nextgroupplot[title={\small GPT 2.7B}, xlabel={\footnotesize Global Batch Size},]
\addplot+[mark size=1.pt, mark=diamond*] coordinates {(16,2.55) (32,3.35) (48,3.27)};
\addlegendentry{FSDP}
\addplot+[mark size=1.pt, mark=triangle*] coordinates {(16,2.55) (32,3.43) (48,4.07) (64,3.68)};
\addlegendentry{\systemns-CB}
\addplot+[mark size=1.pt, mark=x] coordinates {(16,1.48) (32,2.21) (48,2.33) (64,2.48)
(80,2.57) (96,2.58) (112,2.54)
(128,2.61) (144,2.53) (160,2.61) (176,2.63)
(192,2.62) (208,2.63) (224,2.63) (240,2.55) (256,2.64)};
\addlegendentry{\systemns-MB}
\addplot+[mark size=1.pt,mark=*] coordinates {(16,2.65) (32,4.2) (48,4.18) (64,4.45) 
(80,4.51) (96,4.62) (112,4.61) (128,4.57) (144,4.59) (160,4.52) (176,4.46) 
(192,4.55) (208,4.6) (224,4.61) (240,4.62) (256,4.67)};
\addlegendentry{\system}

\node at (axis cs:64,3.68) [anchor=west] {\tiny\textcolor{red}{\textit{OOM}}};
\node at (axis cs:48,3.27) [anchor=west] {\tiny\textcolor{blue}{\textit{OOM}}};

\nextgroupplot[title={\small Llama 3B}]
\addplot+[mark size=1.pt, mark=diamond*] coordinates {(16,2.04) (32,2.96) (48,3.14)};
\addplot+[mark size=1.pt, mark=triangle*] coordinates {(16,2.02) (32,3.47) (48,3.6)};
\addplot+[mark size=1.pt, mark=x] coordinates {(16,1.22) (32,1.87) (48,2.42) (64,2.75)
(80,2.9) (96,2.92) (112,2.93) (128,2.95) (144,2.96) (160,2.95) 
  (176,2.97) (192,2.97) (208,2.97) (224,2.97) (240,2.98) (256,2.97)};
\addplot+[mark size=1.pt, mark=*] coordinates {(16,2.43) (32,3.99) (48,4.3) (64,4.33) 
(80,4.48) (96,4.6) (112,4.57) (128,4.67) (144,4.59) (160,4.77) (176,4.71) 
(192,4.65) (208,4.58) (224,4.64) (240,4.67) (256,4.85)};

\node at (axis cs:48,3.7) [anchor=west] {\tiny\textcolor{red}{\textit{OOM}}};
\node at (axis cs:48,3.24) [anchor=west] {\tiny\textcolor{blue}{\textit{OOM}}};
    
  \end{groupplot}

  \end{tikzpicture}
  \caption{Throughput comparison at different batch sizes
  for \system with, and without, compute and memory balancing.}
  \label{fig:throughput_vs_batch_ablation}
\end{figure}
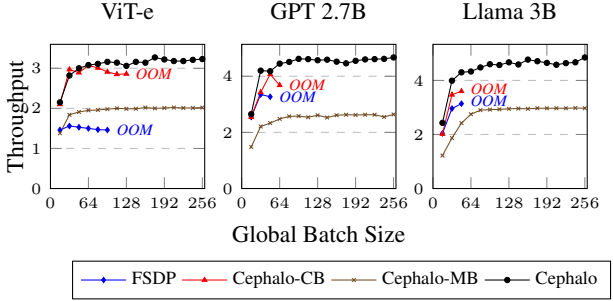

We conducted an ablation study to assess the individual and joint contributions of compute and memory balancing to \systemns's performance. We compared \systemns's training throughput with two variants: compute balancing only (\systemns-CB) and memory balancing only (\systemns-MB), alongside baseline FSDP. Experiments were run on Cluster A with ViT-e, GPT-2.7B, and Llama-3B, scaling batch sizes to 256, as shown in Figure \ref{fig:throughput_vs_batch_ablation}.
\systemns-CB improves throughput over FSDP by balancing compute but encounters out-of-memory (OOM) issues beyond a batch size of 100 for all models, with throughput declining as it nears max memory capacity. \systemns-MB prevents OOM by balancing memory with uneven training state partitioning and using gradient accumulation with a microbatch size of 1. However, its throughput is lower than FSDP’s, as gradient accumulation with such a small microbatch size fails to fully utilize GPU compute, underscoring the need for prudently configuring gradient accumulation.
\system overcomes \systemns-CB and \systemns-MB limitations by jointly balancing compute, memory, and gradient accumulation, essential for high throughput on heterogeneous GPU clusters. It achieves the highest training throughput across all batch sizes and sustains high throughput up to a batch size of 256 without running OOM.

\subsection{Gradient Accumulation Optimizations}
\label{sec:grad_accum}
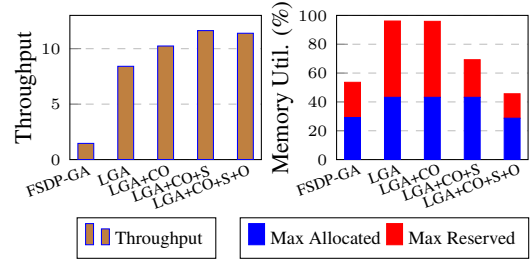
\begin{figure}[]
\centering
\begin{tikzpicture}
    \begin{groupplot}[
        group style={
            group size=2 by 1,
            horizontal sep=10mm
        },
        width=0.5\columnwidth, 
        tick style={tick pos=left},
        height=3.5cm,
        ylabel style={shift={(0pt,-5pt)}}
    ]
    \nextgroupplot[
        ybar, 
        symbolic x coords={FSDP-GA, LGA, LGA+CO, LGA+CO+S, LGA+CO+S+O},
        xtick=data,
        ylabel={\footnotesize Throughput},
        x tick style={draw=none},
        xticklabel style={rotate=20, anchor=east, font=\tiny, xshift=5},
        y tick label style={font=\tiny},
        nodes near coords align={vertical},
        ymin=0, ymax=13, bar width=6pt,
        ymajorgrids=true,
        grid style=dashed,
        legend style={at={(0.4,-0.4)},anchor=north,legend columns=-1, font=\scriptsize},
    ]
    \addplot+[fill=brown] coordinates {(FSDP-GA, 1.46) (LGA, 8.41) (LGA+CO, 10.24) (LGA+CO+S, 11.63) (LGA+CO+S+O, 11.39)};
    \addlegendentry{Throughput}

    \nextgroupplot[
        ybar stacked,
        symbolic x coords={FSDP-GA, LGA, LGA+CO, LGA+CO+S, LGA+CO+S+O},
        xtick=data,
        ylabel={\footnotesize Memory Util. (\%)},
        xticklabel style={rotate=20, anchor=east, font=\tiny, yshift=-4pt, xshift=5},
        y tick label style={font=\tiny},
        x tick style={draw=none},
        legend style={at={(0.5,-0.3)}, anchor=north, legend columns=-1, font=\scriptsize},
        ymin=0, ymax=100,
        , bar width=6pt,
        ymajorgrids=true,
        grid style=dashed,
        legend style={at={(0.25,-0.4)},anchor=north,legend columns=-1, font=\scriptsize},
    ]
    \addplot+[ybar, fill=blue] coordinates {(FSDP-GA, 29.16) (LGA, 43.17) (LGA+CO, 43.17) (LGA+CO+S, 43.17) (LGA+CO+S+O, 28.68)};
    \addlegendentry{Max Allocated}
    \addplot+[ybar, fill=red] coordinates {(FSDP-GA, 24.43) (LGA, 52.92) (LGA+CO, 52.70) (LGA+CO+S, 26.09) (LGA+CO+S+O, 16.96)};
    \addlegendentry{Max Reserved}
    \end{groupplot}
\end{tikzpicture}

\caption{Speedup and memory reduction from
our gradient accumulation optimizations (LGA+CO+S+O) on GPT 6.7B.}
\label{fig:ga_optimizations}
\end{figure}

In Figure \ref{fig:ga_optimizations}, we investigate the throughput and memory improvements obtained from \systemns's gradient accumulation optimizations. Starting from the 
existing gradient accumulation in FSDP (FSDP-GA), we introduce layered gradient accumulation (LGA),
then add communication overlap with computation (CO), compute synchronization (S), and activation
offloading (O). 
We train the 
GPT 6.7B model with a batch size of $256$ (16 microbatches of size 1 per GPU). A homogeneous cluster
of 16$\times$V100 GPUs is used to isolate from the effects of heterogeneous GPUs. 

While FSDP-GA encounters communication bottlenecks, LGA achieves a 6$\times$ speedup by minimizing communication overhead and increases throughput by 22\% through full communication overlap with gradient accumulation. Additionally, compute synchronization and activation offloading eliminate memory overhead and fragmentation, boosting throughput by an extra 11\%. The final implementation with all optimizations (LGA+CO+S+O) delivers a 7.8$\times$ speedup over FSDP-GA while reducing memory usage.

\subsection{Optimized Training Configurations}
\label{sec:opt_config}
\begin{figure}[htbp]
  \centering
  \includegraphics[width=\columnwidth]{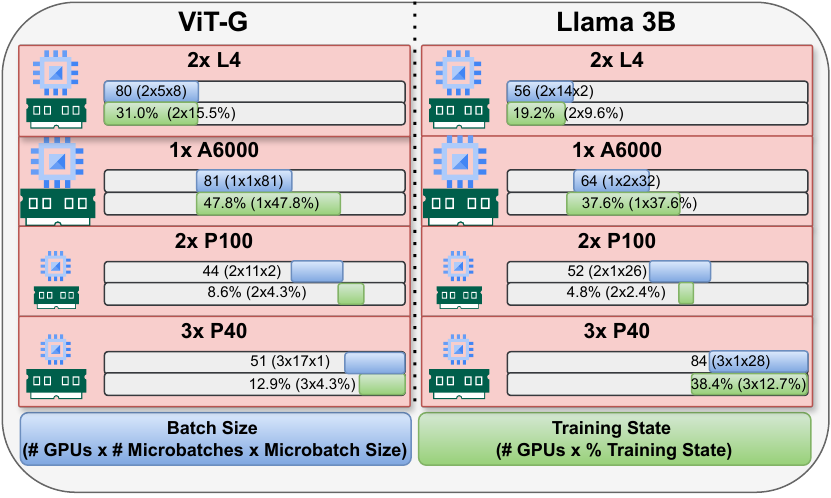}
  \caption{Optimized training configuration for ViT-G \& Llama 3B.}
  \label{fig:opt_config}
\end{figure}
In Figure \ref{fig:opt_config}, we show \systemns's optimized configurations for ViT-G and Llama 3B on Cluster A with batch size 256. The A6000 GPU, being faster and having more memory than the L4s, P100s, and P40s, is assigned the largest portion of the training state and compute. The L4s, with about half the compute and memory of the A6000, receive roughly half the batch size and training state. P100s and P40s are assigned smaller batch sizes, with the P40 handling a larger training state due to its greater memory capacity.

\section{Conclusion}
\system is the first system that jointly resolves imbalances in compute and memory 
across GPUs when training on a heterogeneous cluster. 
It decouples compute and memory requirements for each GPU through uneven compute division, parameter sharding, and gradient accumulation. \system models compute, memory, and communication holistically and uses an optimizer to optimally allocate training state, batch size, and gradient accumulation across GPUs. Evaluations on multiple clusters show that \system achieves significantly higher training throughput while supporting larger
models and batch sizes than existing systems.

\bibliography{main_arxiv}
\bibliographystyle{main_arxiv}

\clearpage
\appendix
\section{Supplementary Material for Optimizer}

\subsection{Dynamic Programming Algorithm}
Algorithm \ref{alg:dp} gives the pseudocode for the dynamic programming algorithm used by \systemns's optimizer
to determine an optimal assignment of batch sizes, gradient accumulation, and training state to each GPU. Notation is defined in Table \ref{table:notation_copy}.

\begin{table}[h!]
\centering
\resizebox{\columnwidth}{!}{%
\begin{tabular}{|c|l|}
\hline
\textbf{Symbol} & \textbf{Description} \\ \hline
$N, B$           & Number of layers and GPUs \\ \hline
$m_i, \ell_i$, $g_i$   & Microbatch size and number of microbatches for $i$th GPU $g_i$ \\ \hline
$M(m)$          & Compute memory for microbatch size $m$ \\ \hline
$M_{cap}^{g_i}$ & Memory capacity of $g_i$ \\ \hline
$T^{g_i}_f(m, \ell)$ & Forwards latency of $g_i$ for $\ell$ microbatches of size $m$ \\ \hline
$T_b^{g_i}(m, \ell)$ & Backwards latency of $g_i$ for $\ell$ microbatches of size $m$ \\ \hline
$AG$, $RG$           & \textit{AllGather} and \textit{ReduceScatter} latency \\ \hline
$M_{state}^{es}$ & Memory required to store an even training state share \\ \hline
\end{tabular}%
}
\caption{Notation and Definitions}
\label{table:notation_copy}
\end{table}
\begin{algorithm}
  \caption{Throughput Maximization using DP}\label{alg:dp}
  \begin{algorithmic}
  
  \STATE {\bfseries Input:} \# of GPUs $N$, Batch Size $B$
  \STATE {\bfseries Output:} Training configuration $solution$

  \STATE Initialize $D[0 \dots N][0 \dots B][0 \dots B]$ with $\infty$
  \STATE $D[0][0][0] \gets 0$
  
  \FOR{$i \gets 1$ {\bfseries to} $N$}
    \FOR{$j \gets 1$ {\bfseries to} $B$}
      \FOR{$k \gets 1$ {\bfseries to} $j$}
        \FOR{$m \gets 1$ {\bfseries to} $k$}
          \FOR{$\ell \gets 1$ {\bfseries to} $\lfloor j / m \rfloor$}
            \IF{$M(m,\ell) > M_{cap}^{g_{i}}$}
              \STATE \textbf{continue} with the next $m$
            \ENDIF
            \STATE $AG' \gets AG$, $RS' \gets RS$
            \IF{$M(m,\ell) + M_{state}^{es} > M_{cap}^{g_{i}}$}
              \STATE $AG' \gets AG_{uneven}$, $RS' \gets RS_{uneven}$
            \ENDIF
            \STATE $T_{i,\ell,m} \gets \max(T_f^{g_i}(m, \ell), AG') + \max(T_b^{g_i}(m, \ell), AG' + RS')$
            \STATE $R \gets \max(D[i-1][j-\ell \cdot m][k-m], T_{i,\ell,m})$
            \STATE $D[i][j][k] \gets \min(D[i][j][k], R)$
          \ENDFOR
        \ENDFOR
      \ENDFOR
    \ENDFOR
  \ENDFOR
  
  \STATE $minimumLatency \gets 0$, $solution \gets \textbf{None}$
  
  \FOR{$k \gets 1$ {\bfseries to} $B$}
      \IF{$D[N][B][k] < minimumLatency$}
          \STATE $minimumLatency \gets D[N][B][k]$
          \STATE $solution \gets \text{Backtrack}(D[N][B][k])$
      \ENDIF
  \ENDFOR

  \end{algorithmic}
\end{algorithm}

\subsection{Optimization Time}
\label{sec:opt_time}
\begin{table}[ht]
  \centering
  \resizebox{0.3\textwidth}{!}{%
  \begin{tabular}{lc}
  \hline
  \textbf{Subtask} & \textbf{Runtime (s)} \\
  \hline
  Profile Compute & 23 \\
  Profile Memory & 486 \\
  Profile Communication & 150 \\
  Partition Compute DP & 327 \\
  Partition State & 1 \\\hline
  \textbf{Total} & \textbf{987} \\
  \hline
  \end{tabular}}
  \caption{Breakdown of profiling and optimization runtime.}
  \label{tab:subtask_runtime}
  \end{table}
To generate a training configuration, \system profiles the model
and network in addition to running the optimizer to partition compute and training state.
This runtime depends on the number of GPUs in the cluster, size of the model, and 
batch size. Even in our largest experiment with 64 GPUs, GPT 6.7B and a batch size
of 512, it took less than 20 minutes to generate the training configuration. The search time 
is negligible compared to the long times required to train these large models.
Moreover, the profiling tasks need to be run only once for a given model and cluster. The optimizer
can reuse the profiling data to generate configurations for different batch sizes and GPUs.
Table \ref{tab:subtask_runtime} shows the runtime breakdown for each subtask in the optimization process.

\subsection{Performance Model Accuracy}
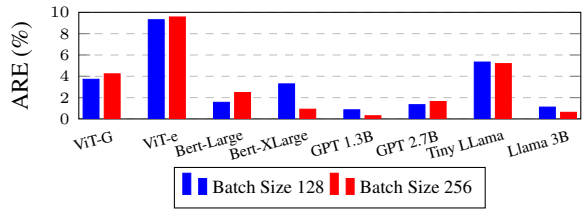
\begin{figure}
\centering
\begin{tikzpicture}
\begin{axis}[
    ybar,
    enlarge x limits=0.05,
    legend style={at={(0.5,-0.15)},
      anchor=north,legend columns=-1},
    ylabel={\footnotesize ARE (\%)},
    symbolic x coords={ViT-G,ViT-e,Bert-Large,Bert-XLarge,GPT 1.3B,GPT 2.7B,Tiny LLama,Llama 3B},
    xtick=data,
    x tick label style={font=\tiny, rotate=15, anchor=east, xshift=7, yshift=-3},
    x tick style={draw=none},
    nodes near coords align={vertical},
    width=\columnwidth, 
    height=3.cm,
    y tick label style={font=\tiny},
    bar width=6pt,
    ymin=0, ymax=10,
    ymajorgrids=true,
        grid style=dashed,
    legend style={at={(0.5,-0.45)},anchor=north,legend columns=-1, font=\scriptsize},
  ]
\addplot+[mark size=1.pt, fill=blue] coordinates {(ViT-G,3.718) (ViT-e,9.305) (Bert-Large,1.5574) (Bert-XLarge,3.2872) (GPT 1.3B,0.8546) (GPT 2.7B,1.3364) (Tiny LLama,5.3348) (Llama 3B,1.1096)};
\addplot+[mark size=1.pt, fill=red] coordinates {(ViT-G,4.2295) (ViT-e,9.5628) (Bert-Large,2.4768) (Bert-XLarge,0.9120) (GPT 1.3B,0.2963) (GPT 2.7B,1.6316) (Tiny LLama,5.1867) (Llama 3B,0.6203)};
\legend{Batch Size 128,Batch Size 256}
\end{axis}
\end{tikzpicture}
\caption{Performance model absolute relative  error (ARE).}
\label{fig:latency_model_error}
\end{figure}
\systemns's optimizer uses a performance model to predict runtime across training configurations, which is essential for efficiently navigating the large search space and optimizing configurations. Figure \ref{fig:latency_model_error} shows the absolute relative error between predicted and actual latencies on Cluster A. Across all models and batch sizes, errors remained within 10\%, with a mean absolute relative error of 2.9\%. Notably, error rates did not increase for larger models or batch sizes, demonstrating the model's robustness.

\begin{figure*}[]
  \centering
  \includegraphics[width=\textwidth]{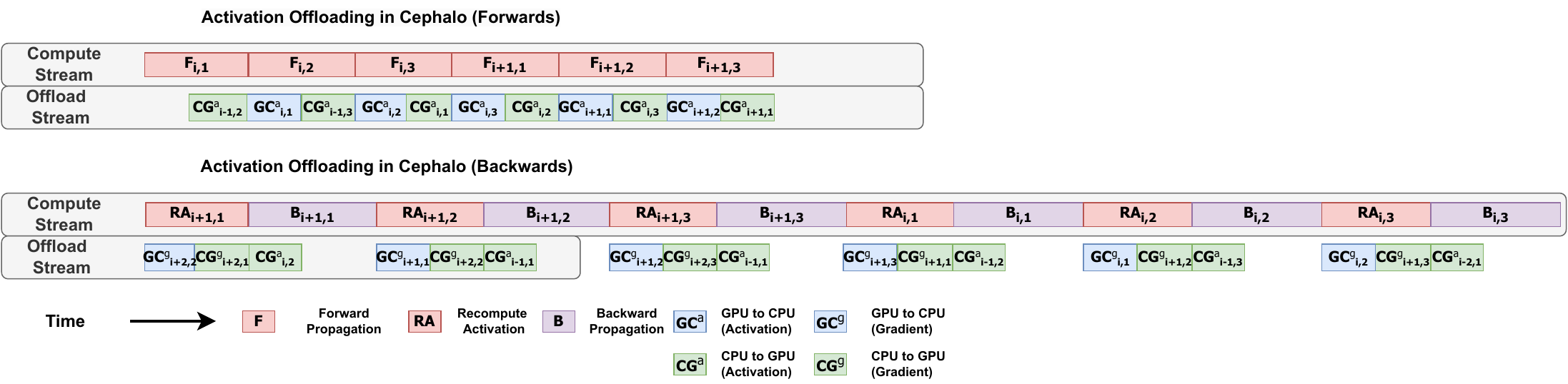}
  \caption{Activation Offloading in Layered Gradient Accumulation. We visualize the sequence of offloading to perform forwards and backwards for two consecutive model layers, $i, i+1$ on 3 microbatches. It assumes gradient checkpointing, recomputing activations in the backwards pass (RA). $GC^a_{ij}$ refers to moving the activation computed by the $i$th layer for the $j$th microbatch from GPU to CPU. $CG^a_{ij}$ refers to moving the same value from CPU to GPU. $GC^g_{ij}$ corresponds to moving the gradient of the activation produced by the $i$th layer for the $j$th microbatch from GPU to CPU. Finally, $CG^g_{ij}$ corresponds to moving that same value from CPU to GPU.}
  \label{fig:activation_offload}
\end{figure*}
\section{Supplementary Material for Activation Offloading}
In layered gradient accumulation, we implement activation offloading such that
we avoid excessive memory overheads from holding activations through
multiple microbatches of communication. Our implementation executes the offloading 
on a separate stream so it does not block computation. When the activations are needed
again, they are also prefetched to overlap with computation. We visualize this process 
for gradient accumulation with $3$ microbatches in Figure \ref{fig:activation_offload}.

During the forwards:
\begin{enumerate}
    \item After the activation is computed for the current microbatch, it is offloaded to the CPU while the next microbatch runs.
    \item Before the next microbatch runs, we prefetch its input activation from the CPU and overlap it with the execution of the current microbatch.
\end{enumerate}

During the backwards pass:
\begin{enumerate}
    \item After the gradient is computed for the current microbatch, we offload the gradient to CPU.
    \item Before the activations are recomputed for the next microbatch, we prefetch its input activation from CPU.
    \item Before the gradients are computed for the next microbatch, we prefetch the gradient of the previous layer from the GPU, which is needed to compute
    the gradient.
\end{enumerate}

\section{Supplementary Material For Communication Latency}
\begin{figure}[htbp]
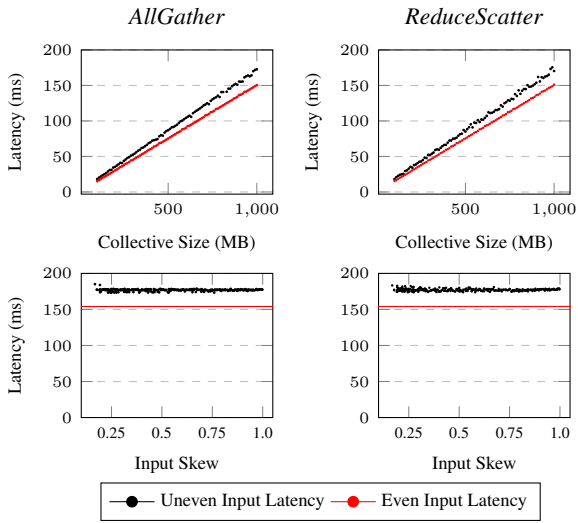

  \centering
  \begin{tikzpicture}
    \begin{groupplot}[
      group style={
        group size=2 by 2,
        horizontal sep=40pt,
        vertical sep=30pt,
      },
      width=0.5\linewidth,
      legend pos=outer north east,
      legend style={
            font=\scriptsize, 
            at={(0.1,-0.45)}, 
            anchor=north west,
            legend columns=-1, 
        },
      grid style=dashed,
      ymajorgrids=true
    ]
    
    \nextgroupplot[
      title={\small\textit{AllGather}},
      xlabel={\scriptsize Collective Size (MB)},
      y tick label style={font=\tiny},
      x tick label style={font=\tiny},
      ylabel={\scriptsize Latency (ms)},
      scatter src=explicit symbolic,
      scatter/classes={
        a={mark=*,black}
      },
      ymax=200,
      height=3.5cm,
    ]

    
    \addplot[forget plot, scatter,only marks,scatter src=explicit symbolic, mark size=0.3pt, color=black]
      coordinates {
(100, 18.10803125)(110, 20.204749999999994)(120, 21.711312499999995)(130, 23.470437500000006)(140, 24.95065625)(150, 26.478906249999998)(160, 28.801999999999996)(170, 30.82303125)(180, 31.157999999999998)(190, 34.337406249999994)(200, 35.23646875000001)(210, 37.19015625)(220, 38.85596875)(230, 41.06846875)(240, 40.64109374999999)(250, 43.74553125)(260, 45.65771874999999)(270, 47.8273125)(280, 49.06606249999999)(290, 51.167874999999995)(300, 52.452875000000006)(310, 53.899375)(320, 55.822781250000006)(330, 58.05687500000001)(340, 59.252531250000025)(350, 61.80809374999999)(360, 62.716093749999985)(370, 64.44415624999999)(380, 66.3128125)(390, 68.66546875)(400, 69.90521874999999)(410, 72.23028125)(420, 72.84559374999999)(430, 73.66528124999999)(440, 74.405)(450, 78.34662500000002)(460, 79.94856249999998)(470, 79.27968749999998)(480, 83.59384374999999)(490, 85.81956249999998)(500, 86.85418749999998)(510, 88.67624999999998)(520, 90.92531249999996)(530, 91.45640625)(540, 93.03790625000002)(550, 94.69528125)(560, 97.09578124999999)(570, 99.45124999999999)(580, 100.63706249999998)(590, 102.4208125)(600, 104.29725000000002)(610, 105.88112499999998)(620, 107.34287499999999)(630, 105.73534375000001)(640, 110.85165625000005)(650, 112.81521875)(660, 114.4728125)(670, 116.07290625)(680, 117.66168750000003)(690, 119.66024999999999)(700, 121.85109375000002)(710, 123.17709374999998)(720, 124.90049999999998)(730, 123.58468750000003)(740, 128.290125)(750, 130.40081249999997)(760, 131.99753125)(770, 133.29268749999997)(780, 135.61778125)(790, 133.70118750000003)(800, 140.588)(810, 140.72162500000002)(830, 141.15531250000004)(840, 145.47881250000003)(850, 145.58621874999994)(860, 148.79515624999996)(870, 150.70781250000002)(880, 152.17118749999997)(890, 151.02043749999996)(900, 155.50243749999998)(910, 155.22443750000002)(920, 155.54546875)(930, 161.3729375)(940, 162.86259375)(950, 164.05624999999998)(960, 162.81306249999997)(970, 164.22537499999996)(980, 170.07518749999997)(990, 171.91275000000002)(1000, 172.70040625000001)
      };

      \addplot[forget plot, color=red, solid, scatter,only marks,scatter src=explicit symbolic, mark size=0.3pt] coordinates {
        (100, 15.3895625)(110, 16.548093750000003)(120, 18.077562500000003)(130, 20.051593750000002)(140, 21.094281249999998)(150, 22.57846875)(160, 24.12431250000001)(170, 25.591687499999995)(180, 27.083687499999996)(190, 28.69478125000001)(200, 30.190124999999995)(210, 31.5799375)(220, 33.1645)(230, 35.000843749999994)(240, 36.09628124999999)(250, 37.638437499999995)(260, 39.695343750000006)(270, 40.624906249999995)(280, 42.13315625)(290, 44.244843749999994)(300, 45.123125)(310, 46.639125)(320, 48.18175000000001)(330, 49.63146875)(340, 51.12781250000001)(350, 52.785531250000005)(360, 54.14634374999999)(370, 55.64931250000001)(380, 57.18178125)(390, 59.20053125000001)(400, 60.193593750000005)(410, 61.78059375000001)(420, 63.89565625000001)(430, 64.70653125)(440, 66.23653125000001)(450, 68.52034375)(460, 69.24046875)(470, 70.68871874999999)(480, 72.23765625)(490, 73.75525)(500, 75.23171875)(510, 76.86784374999999)(520, 78.33328125000001)(530, 79.8455625)(540, 81.23334375000002)(550, 83.20474999999999)(560, 84.24937499999999)(570, 85.74093750000002)(580, 88.2198125)(590, 88.7836875)(600, 90.29437500000002)(610, 92.80431250000002)(620, 93.33909375)(630, 94.77709375)(640, 96.3329375)(650, 97.92062500000002)(660, 99.25815625)(670, 100.77571875000001)(680, 102.3799375)(690, 103.8589375)(700, 105.28643750000002)(710, 107.71396874999998)(720, 108.3965625)(730, 109.87228125000001)(740, 112.37390625)(750, 112.85981250000002)(760, 114.42634375000002)(770, 116.93290624999997)(780, 117.40728124999997)(790, 119.04953125000002)(800, 120.42831249999999)(810, 121.94331250000002)(820, 123.38568750000002)(830, 125.14865625)(840, 126.43800000000002)(850, 128.01006250000003)(860, 129.3473125)(870, 131.5425)(880, 132.44078125000001)(890, 134.03187499999996)(900, 136.49193749999998)(910, 136.93996875000002)(920, 138.39278125)(930, 140.8859375)(940, 141.405875)(950, 142.8386875)(960, 144.38225000000006)(970, 145.922875)(980, 147.30896875)(990, 148.98371875)(1000, 150.36474999999996)
      };
      
    \nextgroupplot[
      title={\small\textit{ReduceScatter}},
      xlabel={\scriptsize Collective Size (MB)},
      ylabel={\scriptsize Latency (ms)},
       scatter src=explicit symbolic,
       y tick label style={font=\tiny},
      x tick label style={font=\tiny},
      scatter/classes={
        a={mark=*,black}
      },
      ymax=200,
      height=3.5cm,
    ]
    
    \addplot[scatter,only marks,scatter src=explicit symbolic, mark size=0.3pt,]
      coordinates {
        (100, 18.196406249999995)(110, 20.803062499999996)(120, 22.686781250000003)(130, 24.173656249999997)(140, 24.924718750000004)(150, 25.507156249999994)(160, 29.574500000000004)(170, 30.437375000000007)(180, 32.842218749999994)(190, 35.39484375000001)(200, 34.30203125)(210, 37.81796875)(220, 38.74490625)(230, 40.58375)(240, 42.56340625000001)(250, 42.983437499999994)(260, 45.100750000000005)(270, 48.38437500000002)(280, 47.83559374999999)(290, 52.32575)(300, 52.21740625)(310, 53.72646875000001)(320, 55.511281249999996)(330, 59.12481249999998)(340, 60.037593750000006)(350, 59.427656250000005)(360, 62.163875)(370, 64.44909375)(380, 66.23021875000002)(390, 66.44659374999999)(400, 69.46796875)(410, 72.39778125)(420, 72.52734374999999)(430, 75.06684374999999)(440, 75.22921875)(450, 76.46974999999998)(460, 79.17243750000002)(470, 81.40043749999998)(480, 82.06903125)(490, 87.20390625000002)(500, 85.03843750000001)(510, 88.32025000000002)(520, 91.06546874999998)(530, 95.27406249999999)(540, 90.944625)(550, 97.66159375)(560, 97.11737500000002)(570, 101.17221875)(580, 98.52421875)(590, 102.11874999999999)(600, 101.82559374999998)(610, 104.41756250000002)(620, 107.12921874999999)(630, 112.22549999999998)(640, 111.14284375000003)(650, 111.49043750000001)(660, 112.97368750000001)(670, 115.37550000000003)(680, 115.36371875)(690, 116.78965625)(700, 122.03596875)(710, 120.18509375)(720, 124.96503125000001)(730, 126.43396875000002)(740, 129.064125)(750, 128.51878125000005)(760, 131.217)(770, 130.7045625)(780, 137.80171875)(790, 135.24112499999998)(800, 143.11537500000003)(810, 138.22521874999995)(820, 142.63390625)(830, 145.0315)(840, 144.89912499999997)(850, 150.38115624999998)(860, 148.13690624999998)(870, 148.44234375000002)(880, 149.71162500000005)(890, 151.61509375)(900, 153.86899999999997)(910, 161.897875)(920, 155.68121875000003)(930, 161.62834375000003)(940, 160.54559375000005)(950, 162.5505625)(960, 163.26803125)(970, 165.20690625000003)(980, 173.22512499999996)(990, 175.50590625000004)(1000, 170.35196875000003)
      };

      \addplot[color=red, solid, scatter,only marks,scatter src=explicit symbolic, mark size=0.25pt] coordinates {
      (100, 15.435343750000005)(110, 16.586468749999998)(120, 18.2350625)(130, 20.108)(140, 21.124312500000006)(150, 22.625437500000007)(160, 24.268156250000004)(170, 25.763906249999994)(180, 27.166062499999995)(190, 28.788125000000004)(200, 30.42190625)(210, 31.899156249999994)(220, 33.28253124999999)(230, 35.16203124999999)(240, 36.273781250000006)(250, 37.6826875)(260, 39.78156249999999)(270, 40.802125000000004)(280, 42.23443749999999)(290, 44.45453125)(300, 45.35965624999999)(310, 46.764999999999986)(320, 48.318968749999996)(330, 49.91187500000001)(340, 51.22059374999999)(350, 52.76440624999998)(360, 54.70937500000001)(370, 55.81815625)(380, 57.2361875)(390, 59.4015625)(400, 60.34375)(410, 61.85856250000001)(420, 64.01715625)(430, 64.89753124999999)(440, 66.27940625)(450, 68.6120625)(460, 69.44678124999999)(470, 70.8715)(480, 72.41665624999999)(490, 74.07646875)(500, 75.31628125000002)(510, 76.80075000000001)(520, 78.74375000000002)(530, 79.86178124999998)(540, 81.3019375)(550, 83.56543749999999)(560, 84.29653125000002)(570, 85.81934375)(580, 88.22065625000002)(590, 88.91381249999999)(600, 90.422875)(610, 92.83140625)(620, 93.77015625)(630, 94.83603125)(640, 96.41328125)(650, 97.95128124999998)(660, 99.29384375000001)(670, 100.853125)(680, 102.5595625)(690, 103.95356250000002)(700, 105.41878125)(710, 107.70343750000002)(720, 108.28840625)(730, 109.91709375)(740, 112.33593750000003)(750, 112.9704375)(760, 114.37724999999999)(770, 116.91234374999999)(780, 117.33171875000002)(790, 118.83787499999997)(800, 120.38146875000001)(810, 121.99437499999999)(820, 123.40240624999998)(830, 124.88565625)(840, 126.52159375000002)(850, 127.83637500000002)(860, 129.40403125000003)(870, 131.80487500000004)(880, 132.36384375)(890, 133.9439375)(900, 136.40303125)(910, 136.96084374999998)(920, 138.55793750000004)(930, 141.05381250000002)(940, 141.8520625)(950, 142.93778125)(960, 144.62328125000002)(970, 146.58990624999998)(980, 147.5973125)(990, 148.932)(1000, 151.20765624999999)
      };

    \nextgroupplot[
      xlabel={\scriptsize Input Skew},
      ylabel={\scriptsize Latency (ms)},
      y tick label style={font=\tiny},
      x tick label style={font=\tiny},
      scatter src=explicit symbolic,
      scatter/classes={
        a={mark=*,black}
      },
      xmin=0.1,
      xmax=1.05,
      ymin=0,
      ymax=200,
       xtick={0.25,0.5,0.75,1.0},
      xticklabels={0.25,0.5,0.75,1.0},
      height=3.5cm,
    ]
    \addlegendimage{mark=*, color=black, mark size=2pt}
    \addlegendentry{Uneven Input Latency};
    \input{allgather_coordinates.tex}

    \addlegendimage{mark=*, color=red, mark size=2pt}
    \addlegendentry{Even Input Latency};

    \addplot[forget plot, color=red, solid] coordinates {(0,154.093625) (2,154.093625)};

    \nextgroupplot[
      xlabel={\scriptsize Input Skew},
      y tick label style={font=\tiny},
      x tick label style={font=\tiny},
      scatter src=explicit symbolic,
      scatter/classes={
        a={mark=o,draw=red}
      },
      ymin=0,
      ymax=200,
      xmin=0.1,
      xmax=1.05,
       xtick={0.25,0.5,0.75,1.0},
      xticklabels={0.25,0.5,0.75,1.0},
      height=3.5cm,
    ]
    \input{reducescatter_coordinates}
        \addplot[forget plot, color=red, solid] coordinates {(0,154.15121875) (2,154.15121875)};

    \end{groupplot}
  \end{tikzpicture}
  \caption{NCCL collective latencies
for uneven vs even sized inputs for different collective sizes (top), and input skew (bottom). 
Latencies were profiled on a heterogeneous $8$-GPU
cluster (Cluster A, Section \ref{sec:experimental_setup}).}
  \label{fig:latency_plots_scale}
\end{figure}
Let the collective size be the sum of the input sizes to an \textit{AllGather} or \textit{ReduceScatter} 
collective. We made two general observations from analyzing the communication latencies in relation to the
collective size with randomly generated input sizes versus even input sizes:
\begin{enumerate}
    \item There is a strong correlation between communication latency and collective size for 
both uneven and even inputs, as shown in Figure \ref{fig:latency_plots_scale} where we plot collective latency against input size.
    \item Communication latency remains consistent across varying input sizes for a given collective size, 
defined by the degree of input skew—the ratio of the largest input to the total input size. 
Figure \ref{fig:latency_plots_scale} illustrates that latency stays within a narrow range, regardless of input skew.
\end{enumerate}
Based on these observations, we profile the collective latency for evenly sharded training state, which is constant for all layers.
We then assume a conservative $15\%$ overhead in 
communication latency when unevenly sharding the training state for both \textit{AllGather} and \textit{ReduceScatter}. 
\begin{table*}[ht!]
\centering
\caption{Throughput comparison of different models and batch
 sizes on 8-GPU Cluster A. \textit{OOM} denotes Out-of-Memory.}
\label{table:model_strategy_performance_extra}
\renewcommand{\arraystretch}{0.8}
\resizebox{\textwidth}{!}{
\begin{tabular}{@{}lrrrrrrrrrrrrrrrr@{}}
\toprule
\textbf{System} & \multicolumn{2}{c}{\textbf{ViT-G}} & \multicolumn{2}{c}{\textbf{ViT-e}} & \multicolumn{2}{c}{\textbf{Bert-Large}} & \multicolumn{2}{c}{\textbf{Bert-XLarge}} & \multicolumn{2}{c}{\textbf{GPT 1.3B}} & \multicolumn{2}{c}{\textbf{GPT 2.7B}} & \multicolumn{2}{c}{\textbf{Tiny Llama}} & \multicolumn{2}{c}{\textbf{Llama 3B}} \\ 
\cmidrule(lr){2-3} \cmidrule(lr){4-5} \cmidrule(lr){6-7} \cmidrule(lr){8-9} \cmidrule(lr){10-11} \cmidrule(lr){12-13} \cmidrule(lr){14-15} \cmidrule(lr){16-17}
& \textbf{128} & \textbf{256} & \textbf{128} & \textbf{256} & \textbf{128} & \textbf{256} & \textbf{128} & \textbf{256} & \textbf{128} & \textbf{256} & \textbf{128} & \textbf{256} & \textbf{128} & \textbf{256} & \textbf{128} & \textbf{256} \\ 
\midrule
FSDP       & 3.92 & \textit{OOM} & \textit{OOM} & \textit{OOM} & 24.50 & 28.24 & 7.06  & \textit{OOM} & \textit{OOM} & \textit{OOM} & \textit{OOM} & \textit{OOM} & 10.62 & \textit{OOM} & \textit{OOM} & \textit{OOM} \\
Whale      & \textit{OOM} & \textit{OOM} & \textit{OOM} & \textit{OOM} & 27.13 & 28.84 & \textit{OOM} & \textit{OOM} & \textit{OOM} & \textit{OOM} & \textit{OOM} & \textit{OOM} & \textit{OOM} & \textit{OOM} & \textit{OOM} & \textit{OOM} \\
HAP        & \textit{OOM} & \textit{OOM} & \textit{OOM} & \textit{OOM} & 17.48 & 18.54 & \textit{OOM} & \textit{OOM} & \textit{OOM} & \textit{OOM} & \textit{OOM} & \textit{OOM} & \textit{OOM} & \textit{OOM} & \textit{OOM} & \textit{OOM} \\
\system     & \textbf{6.38} & \textbf{6.41} & \textbf{3.02} & \textbf{3.23} & \textbf{33.55} & \textbf{33.69} & \textbf{11.47} & \textbf{11.72} & \textbf{6.83} & \textbf{7.09} & \textbf{4.57} & \textbf{4.67} & \textbf{12.58} & \textbf{12.91} & \textbf{4.51} & \textbf{4.85} \\

\bottomrule
\end{tabular}
}
\end{table*}

\section{Supplementary Material For Evaluation}
\subsection{Additional Baselines}
We also compared \system to:
\begin{itemize}
  \item Whale \cite{whale}: Balances computational loads with data parallelism by assigning batch 
  sizes to GPUs based on their runtime profiles.
  \item HAP \cite{hap}: Uses tensor parallelism across nodes and data parallelism within nodes. 
  The batch size and parameters are sharded unevenly to balance the workload.
\end{itemize}

\subsection{Additional Experiments}
Table \ref{table:model_strategy_performance_extra} compares the trianing throughput of \system to the additional baselines.

\textbf{\textit{Comparison to Whale.}}
~Like \systemns, Whale optimizes compute utilization in the cluster by assigning varying local batch sizes 
to GPUs based on their compute capabilities. However, it is able to train only the smallest model, 
Bert-Large, without running out of memory.
In this cluster, although P40 GPUs have similar compute speeds to P100s, they have twice 
the memory (24 GB). Despite this, to maintain compute balance, Whale assigns similar batch 
sizes to both, causing P100s to run out of memory when P40s have utilized only 50\% of their memory. 
\system avoids this issue by partitioning compute independently from memory. It assigns a similar batch size
for both P40 and P100 GPUs, but stores a larger share of the training state in the P40 GPUs to balance memory utilization.
Whale also consumes considerably more memory than \system since data parallelism replicates the entire training state
across each GPU. \system saves memory by sharding the training state at the cost of extra communication. 
However, \system effectively masks this extra communication by overlapping it with computation. 
\textbf{\textit{Comparison to HAP.~}}
HAP, like \systemns, can partition compute by dividing the batch size unevenly across GPUs. However, HAP relies on
tensor parallelism to partition the training state, which is proportional to the amount of compute assigned.
HAP does not consider the memory constraints on the GPUs, so it runs out of memory on all models but Bert-Large.
Despite the compute partitioning, HAP is unable to train efficiently due to the high communication 
overheads of tensor parallelism, which requires high-bandwidth interconnects.

%


\end{document}